\documentclass[prd,superscriptaddress,11pt]{revtex4-1}
\usepackage{amsmath}    
\usepackage{natbib}
\usepackage[english]{babel}
\usepackage[utf8]{inputenc}
\usepackage{graphicx}   
\usepackage{slashed}
\usepackage{epstopdf}
\usepackage{verbatim}   
\usepackage{color}      
\usepackage{subcaption}
\usepackage{multirow}
\usepackage{hyperref}   
\usepackage{float}
\restylefloat{table}
\usepackage{bm}
\newcommand{\be}{\begin{eqnarray}}
	\newcommand{\ee}{\end{eqnarray}}

\newcommand{\M}{{\mathcal M}}

\def\addresses#1#2{\hbox to \hsize{\@tablebox{#1}\hfil\@tablebox{#2}}}
\def\@tablebox#1{\vtop{\hsize=5in \begin{flushleft} #1 \end{flushleft}}}

\def\beq{\begin{equation}}
	\def\eeq{\end{equation}}
\def\bit{\begin{itemize}}
	\def\eit{\end{itemize}}
\def\beqa{\begin{eqnarray}}
	\def\eeqa{\end{eqnarray}}
\def\bray{\begin{array}}
	\def\eray{\end{array}}

\def\n{\nonumber}

\definecolor{orange}{rgb}{1,0.5,0}
\definecolor{blue}{rgb}{0,0,1}

\begin{document}
	\title{Massive Gauge Theories from Consistency Conditions of  Amplitudes}
	
	\author{Junmou Chen}
	\email{chenjm@jnu.edu.cn}
	\affiliation{Department of Physics, School of Science and Technology, Jinan University, Guangzhou,  China}
	

	\begin{abstract}
		
		Based on the general principles of Lorentz symmetry and unitarity, we introduce two consistency conditions  -- on-shell gauge symmetry and strong massive-massless continuation  -- in constructing  amplitudes of massive gauge theory with elementary particles.  In particular we argue that  on-shell gauge symmetry can be understood as a consequence of Lorentz symmetry, through mixture of a vector boson and  a scalar with  degenerate mass spectrum. Based on the two conditions, 
combined with the  little group transformation and consistent factorization,  we construct three-point and four-point vector boson/scalar amplitudes, then analyze the underlying physical models.   Given the particle masses, almost all possible vertices, including those involving Goldstone modes, are uniquely fixed. The only exceptions are triple and quartic scalar self-couplings, as well as mixing angles between vacuum expectation values (VEVs) and
scalars. In addition, all particle masses must have the same physical origin. If the  number of vector bosons is smaller than 3, the underlying theories for the amplitudes are either massive gauge theories with spontaneous symmetry breaking (SSB) or Stueckelberg theory. The necessary condition for the latter is that the scalars have equal masses.     We also discuss different models depending on the number of scalars involved.  If the number of vector bosons is larger than 3, the underlying theory must be Yang-Mills theory with SSB.   In both abelian and non-abelian cases,  the specific shape of the Higgs potential cannot be determined, which explains the fact that  scalar self-couplings are undetermined, and the relations between the masses are generally  not linear.

	\end{abstract}
	
	
	\maketitle
	
	\section{Introduction}

	The project of constraining viable gauge theories through Lorentz invariance and analytic structure of S-matrix can be traced back to Weinberg in his work of  \cite{PhysRev.134.B882,PhysRev.135.B1049,WEINBERG1964357}, which generated many beautiful results and physical insights. For example,  by considering S-matrix for a massless spin-1 particle taking part in interactions, Lorentz invariance requires the S-matrix to obey on-shell gauge invariance: $k_{\mu}\M^\mu =0$. Moreover,  by taking the soft limit of the spin-1 particle,  charge conservation can be derived.  Thus an interacting theory of massless spin-1 particles has to be a gauge theory. Similar considerations for massless spin-2 particles lead us to linearized Einstein equation and equivalence principle.  This project got a new life  in the last decade with the development of modern approach of amplitudes, especially  spinor-helicity formalism and Britto-Cachazo-Feng-Witten(BCFW) recursion relation\cite{Britto:2005fq}.  In \cite{Benincasa:2007xk, McGady:2013sga, Schuster:2008nh},  it was found that $U(1)$ little group scaling of three-point amplitudes uniquely fix all possible three-point amplitudes.  Moreover, by  demanding consistent factorization of four-point amplitudes at $s, t, u$ channels when they go to $0$, massless spin-1  and spin-2 interacting theories are uniquely fixed to be Yang-Mills theory and linearized general relativity respectively.  For a fresh review, see also \cite{Arkani-Hamed:2017jhn}.  
	
	Similar projects have also been carried out for massive gauge theory.  As early as 70s,  ref(\cite{Cornwall:1973tb, Cornwall:1974km,  LlewellynSmith:1973yud, Lee:1977yc}) derived gauge invariance and upper bound of the Higgs boson by applying perturbative unitarity   on four-point vector boson scattering amplitudes.   In recent years,  ref(\cite{Arkani-Hamed:2017jhn}) introduced a new framework that generalizes the spinor formalism to  massive amplitudes, bringing new life  to this project.  In it and \cite{Conde_2016} before it,  the authors constructed three-point massive amplitudes, but left four-point untouched.  Following \cite{Arkani-Hamed:2017jhn} there have also been many works that  generalize  this approach  to  four-point amplitudes(\cite{Bachu_2020, Choi:2021szj, Bachu:2023fjn, Liu:2022alx, Lai:2023upa, Christensen:2024xzs}), effective field theory(EFT)(\cite{Dong_2023,Liu_2023,Bresciani:2025toe, Balkin_2022, Durieux_2020}) and others(e.g.\cite{ Balasubramanian_2023,  Wu_2022, Ni:2025xkg, Ema:2024vww, Ballav_2021}).

	Despite those achievements,   the project of constructing massive gauge theories from amplitudes is still not wholly satisfactory.  The spinor formalism, though very useful in constructing three-point amplitudes,  is still essentially making use of unitarity in constructing four-point amplitudes.  Moreover,  the perturbative unitarity approach  itself  neglects some crucial aspects of the underlying physics.  First,    precise cancellation between diagrams from  perturbative unitarity implies  an underlying symmetry.  This symmetry, however, is not manifest in the amplitudes, making the cancellation seems ad hoc.   It's only in the Lagrangian that it's clear  the symmetry is  gauge symmetry with  spontaneous symmetry breaking(SSB). The problem then is how to implement gauge symmetry at the level of amplitudes.  It would be also optimal to trace the origin of massive gauge symmetry to Lorentz symmetry directly like the massless case, instead of extracting it from Lagrangian.  Second, the perturbative unitarity approach proves that an elementary scalar can restore the self-consistency of scattering amplitudes and the related theory. However, it's not the only possible solution.  In  both technicolor theory (\cite{Hill:2002ap, Weinberg:1975gm, Susskind:1978ms}) and composite Higgs models (\cite{Kaplan:1983sm, Dugan:1984hq, Miransky:1988xi, Miransky:1989ds}), unitarity is restored through exchange of many resonances in scattering and ultimately the dissolution of composite particles. This means that we need  a clear criteria on the amplitudes to exclude other possibilities in favor of the elementary particle solution.   Finally,  perturbative unitarity does not apply to all physical scenarios.  In Stueckelberg theory,  amplitudes have no unitarity violating behavior \cite{RUEGG_2004}.  It's more complete  and satisfying to incorporate this case into the project too.

	The purpose of this paper is to find  consistency conditions for amplitudes that solve the above  problems, then apply them  to construct massive amplitudes and the corresponding theories.   For the first problem of gauge symmetry, we notice that gauge symmetry with SSB  manifests itself in Goldstone equivalence theorem   and its precise form the massive Ward identity or on-shell gauge symmetry $k^\mu \mathcal M_\mu = \pm i m_V \mathcal M(\varphi)$ ($\varphi$ is Goldstone boson)\cite{Chanowitz:1985hj, Gounaris:1986cr, Bagger:1989fc}.   For the second problem of selecting elementary particles, we notice an elementary particle interacts point-like, meaning the coupling is constant at tree level.  Based on those analysis we propose the following two conditions in constructing massive amplitudes:
	\begin{itemize}\label{item:consis_cond}
		\item {\bf On-shell gauge symmetry(OGS): the amplitude of  massive vector boson with the polarization vector replaced with $k^\mu$ equals to some corresponding scalar amplitude, i.e.  $k^\mu \mathcal{M}_\mu= \pm i m_V\mathcal{M}(\varphi)$.  The vector boson and the scalar  have a common mass. }
		\item {\bf Strong massive-massless continuation:  the S-matrix remains finite when  one of the particles' mass is taken to be 0 continuously.} 
	\end{itemize}
	The two conditions can be further understood as reflections of the general principle  of massive-massless continuation in different aspects. In particular we will show on-shell gauge symmetry can be derived from   Lorentz symmetry and unitarity with some reasonable assumptions.   In \cite{Chen2025wig}, a construction of particles as unitary and irreducible representations of Poincare that satisfies massive-massless continuation was proposed, providing a solid foundation for this principle.  
	
	After finding the consistency conditions that satisfy massive gauge symmetry and select elementary particles, we proceed to apply those conditions, along with little group transformation and consistent factorization, to construct scattering amplitudes for massive vector bosons and scalars up to four-point. We find our approach to be very powerful: we construct all three-point and four-point amplitudes with different mass combinations.   All couplings, including those related to Goldstone bosons, are uniquely fixed up to an overall factor, with the sole exceptions of triple and quartic scalar self-couplings, as well as mixing angles between vacuum expectation values (VEVs) and
	scalars. Most importantly, we find that  in constructing four-point amplitudes with $VV\rightarrow VV$, at least one scalar $S$ and the subsequent vertices of $VVS$ are needed to make sure gauge symmetry is satisfied. In addition, all particles' masses go to $0$ if the scalar mass goes to $0$, implying they all have the same origin.  Thus we conclude  the only theory that's consistent with the conditions of on-shell gauge symmetry and (strong) massive-massless continuation is Yang-Mills theory with Higgs mechanism, in the general sense that symmetry is spontaneously broken to give masses to all particles.  But the shape of the scalar potential is not specified.  Moreover, we found Higgs theory is not the only possible theory when constructing other amplitudes. For example, the underlying  theory for  two scalars and one vector boson is scalar QED with Stueckelberg mechanism, when the two scalars have equal masses.  In this case, the vector boson obtains mass without ``eating" a dynamic Goldstone mode.

The rest of the paper is organized as follows,  

In Sec. (\ref{sec:condition}) we explain in detail   the  motivations and justifications of the consistent conditions of   on-shell gauge symmetry and strong massive-massless continuation.  

In Sec.(\ref{sec:frame_constrt}) we review and introduce the framework of  constructing amplitudes, including conventions of polarizations vectors, propagators and amplitudes. 

In Sec.(\ref{sec:three-point}), we reconstruct three-point amplitudes for all possible combination of scalars and vector bosons with arbitrary masses.  

In Sec.(\ref{sec:four-point}),  we construct  four-point amplitudes from three-point amplitudes, determine the conditions for it to succeed, discuss the underlying theories and models. 

Finally in Sec.(\ref{sec:conclusion}) we have conclusions and discussions. 


\


\section{Consistency Conditions}
\label{sec:condition}
In this section we will demonstrate in detail that  the two consistency conditions  proposed in Item(\ref{item:consis_cond}) come from combined consideration  of Lorentz symmetry,  the general principle of massive-massless continuation and that all particles involved are elementary.


\subsubsection*{On-shell Gauge Symmetry}

To summarize in advance, we will show  on-shell gauge symmetry $k^
\mu \mathcal M_\mu = im_V\mathcal M(\varphi)$ emerges naturally from  the principles of Lorentz symmetry, unitarity and  massive-massless continuation.


It's known that  particles can be understood as unitary,  irreducible representations of Poincare  group. The vector representation includes both  a spin-1 block and a spin-0 block, which are positive and negative in norms respectively.   In the standard approach,  the spin-0 block with negative norm is eliminated from the spectrum by imposing the condition $k\cdot \epsilon =0$, or equivalently $\partial_\mu V^\mu=0$ in the position space.  Thus only the spin-1 block is taken to be physical.   However, massive vector boson constructed in this way doesn't have a smooth massless limit.  As $m_V\rightarrow 0$,  the longitudinal polarization vector goes as $\epsilon^\mu_L \propto \frac{k^\mu}{m_V}\propto \infty$.   Here we will make an alternative construction that has manifest massive-massless continuation. 

We start with the vector representation  without  constraint, including both the spin-1 and the spin-0 block.  To eliminate the  negative norm of   the spin-0 block, we introduce an  scalar $\varphi$ to mix with the spin-0 block of the vector boson $V^\mu$, which together with $\varphi$ is subject to  the mixing condition of 
	\begin{equation}\label{eq: mix_cond_V-phi}
		\partial^\mu V_\mu = m_V\varphi 
	\end{equation}
Notice because of this condition, the number of  degrees of freedom (d.o.f.) after adding $\varphi$ remains the same as $5-1=4$.   Furthermore, noticing that the left side of Eq.(\ref{eq: mix_cond_V-phi})  only selects  the spin-0 block in $V^\mu$, the  condition simply relates one scalar in the vector representation with one scalar ($\varphi$) in the trivial representation.  So the number of scalars after the mixing condition also remains $1$. 
	
	We can further identify the spin-0 block of $V^\mu$ with $\varphi$ by writing it as $V^\mu_{s=0}= \frac{\partial^\mu \varphi}{m_V}$.   In this case,  Eq.(\ref{eq: mix_cond_V-phi}) becomes the equation of motion for the scalar: $\partial^2 \varphi = m_V^2\varphi$. $\varphi$ is then simply an  auxiliary mode, which purpose is to project out  the scalar in $V^\mu$ from a vector representation to  the trivial (scalar) representation.

Eq.(\ref{eq: mix_cond_V-phi}) serves two purposes at the same time.  First,  it makes sure the negative norm of the spin-0 block is canceled by that of the scalar $\varphi$. Applying the condition on the Hilbert space, it indicates that the spin-0 block ($|V_0\rangle $) must always be combined with the scalar $|\varphi\rangle$ to form a physical state, the norm of which is 
\begin{equation}\label{eq:k_phi_norm}
 (\langle V_0| + \langle \varphi|) (| V_0\rangle + |\varphi\rangle) = (-1)\frac{k^\mu}{m_V}\cdot \frac{k_\mu}{m_V}+1=-1+1=0,
\end{equation}
The Hilbert space is now positive semi-definite, satisfying the requirement of quantum mechanics. 
Second,  the condition enables us to eliminate the $\frac{k^\mu}{m_V}$  term by adding a zero-norm state $| V_0\rangle + |\varphi\rangle$, thus ensuring that  the polarization vectors have smooth massless limits.  


The above conclusions can also be translated  on polarization vectors. Eq.(\ref{eq: mix_cond_V-phi}) in momentum space can be written as 
\begin{eqnarray}\label{eq:phy_con_mass}
	k_\mu \epsilon^\mu_\sigma &=& \pm i m_V \epsilon^\varphi \delta_{\sigma 0},  \ \  \  +\  \text{for incoming} \  \	\  -  \  \text{for outgoing}  \\
	\epsilon^\varphi&=&  \mp i,  \ \  \  -\  \text{for incoming} \  \	\  +\  \text{for outgoing} \nonumber 
\end{eqnarray}
$\epsilon^\varphi$ is the ``polarization" of the scalar $\varphi$.   Eq.(\ref{eq:phy_con_mass}) is also  invariant under 
\begin{equation}\label{eq:gauge_sym_mass}
	\epsilon^\mu\rightarrow \epsilon^\mu + \xi k^\mu/m_V, \ \ \epsilon^\varphi \rightarrow \epsilon^\varphi \mp \xi  i, 
\end{equation} 
meaning the polarization vectors defined by  Eq.(\ref{eq:phy_con_mass}) have a gauge redundancy.  
Moreover, keeping in mind the relative minus sign between the two norms, giving zero norm for the combination of  the shift of $\epsilon_\sigma^\mu$ and the shift of $\epsilon^\varphi$: $(-1)\frac{k^\mu}{m_V}\cdot \frac{k_\mu}{m_V}+1=-1+1=0$ as in Eq.(\ref{eq:k_phi_norm}).

Now let's count the number of degrees of freedom (d.o.f.) in our construction. The vector representation and the scalar $\varphi$ give 5 d.o.f., while Eq.(\ref{eq:phy_con_mass}) and Eq.(\ref{eq:gauge_sym_mass}) give 2 constraints, leaving the final number of d.o.f. to be $5-2=3$, same as the standard construction.  Indeed,   the polarization vectors of the standard construction can be understood as a special case in our construction.  To see this, we can eliminate the scalar $\varphi$ from the polarization vectors by making use of Eq.(\ref{eq:phy_con_mass}).

Our construction has manifest massive-massless continuation. By taking $m_V\rightarrow 0$, our construction reduces to massless spin-1 particle.  First,  the constraint in Eq.(\ref{eq:gauge_sym_mass})(Eq.(\ref{eq: mix_cond_V-phi})) reduces to $k\cdot\epsilon = 0$($\partial^\mu V_\mu =0$). Second,  the symmetry in Eq.(\ref{eq:gauge_sym_mass}) reduces to $\epsilon \rightarrow \epsilon + \lambda k$. Combining the two conditions, we obtain exactly the polarization vectors for massless spin-1 particle.

 Now we are ready to extract the physical implication  of  our construction based on  Eq.(\ref{eq: mix_cond_V-phi}) on amplitudes.    In our construction the amplitude of  massive vector bosons can be written as 
\begin{equation}
	\mathcal M(V) = \epsilon_\sigma^\mu \mathcal M_\mu(V) + \epsilon^\varphi_\sigma \mathcal M(\varphi)
\end{equation}
Polarization vectors are equivalent under the redefinition in Eq.(\ref{eq:gauge_sym_mass}), which means  the gauge redundancy  has no contribution to amplitudes. Therefore we must have
\begin{equation}
	k^\mu \mathcal M_\mu =  \pm i m_V \mathcal M(\varphi)
\end{equation}
This is precisely the   on-shell gauge symmetry in Item(\ref{item:consis_cond}).

Finally,  the kinetic Lagrangian for massive vector boson based on Eq.(\ref{eq: mix_cond_V-phi}) can be derived  by making use of Euler-Lagrangian equations.   The results are
\begin{eqnarray}
	\mathcal L_{V^2} &=& -\frac{1}{4}(\partial_\mu V_\nu -\partial_\nu V_\mu)^2 +\frac{1}{2}m_V^2V_\mu^2 \nonumber \\
	\mathcal L_{V-\varphi}  &=& m_V V_\mu \partial^\mu \varphi \\
	\mathcal L_{\varphi^2} &=&  \frac{1}{2}\partial_\mu \varphi \partial^\mu \varphi \nonumber
\end{eqnarray}
This is exactly the Lagrangian of gauge-Goldstone fields in SSB, with the gauge-Goldstone mixing term $\mathcal L_{V-\varphi}$. Moreover,  the  mass term for the scalar to be zero,  which means it has zero mass without the mixing, thus justifying calling it Goldstone boson.  The details of the derivation are in the appendix \ref{sec:mix_Lagran}.

\subsubsection*{ Strong Massive-Massless Continuation}


On-shell gauge symmetry originates from Lorentz symmetry generally, thus doesn't distinguish between elementary and composite particles, but our purpose is to select theories of elementary particles only. This means that we need  another consistent condition. 

Particles are elementary  in the sense that they remain point-like in interactions at arbitrary energy scale. This is in contrast with composite particles, whose form factors change drastically with increased energy.  When the energy is high enough,  a composite particle would eventually fragment into more elementary components.  From the perspective of effective field theory,  the theory based on the composite particle is only effective up to certain energies.  When the energy increases to a certain point,  the effective theory becomes invalid, there needs to be another theory that's based on more elementary degrees of freedom to replace the former.  In amplitudes, one of the main signals of  the process of effective theory becoming invalid is  the  energy increasing behavior that violates perturbative unitarity.  The fact that amplitudes eventually blow up to infinity as $E\rightarrow \infty$,  reflects the process of the composite particle dissolving into its elementary components.

The above analysis can be captured by the partial wave analysis of amplitudes. However,  
we would like to have  an approach that has the above property  directly reflected in  the analyticity of S-matrix as function of momenta, polarization vectors, as well as masses and couplings. Noticing for longitudinal vector boson, $E\rightarrow \infty$ is equivalent to $m_V\rightarrow 0$ to certain extents.  The condition of unitarity for amplitudes can be expressed as following:   amplitudes remain finite when the vector boson's mass goes to $0$. 
We further generalize the condition from vector boson's mass to any particle's mass in the process. This finally gives  the condition of strong massive-massless continuation in Item (\ref{item:consis_cond}).  In mathematical form, it can be written as 
\begin{equation}
	\lim_{m_i\rightarrow 0 }\mathcal M(\{p_i, \epsilon_i, m_i\}, {\lambda_j}) = \text{finite}
\end{equation}
$i$ being any massive particle.

  Finally we note that, although both strong massive-massless continuation and perturbative unitarity  can be understood as  concret realizations of the physical  differences between elementary and composite particles, the two cannot be guaranteed to be equivalent in every situation.   It's an open question  that can only be determined by the results of construction from the conditions.


\

\

\section{Framework of Constructing Amplitudes}
\label{sec:frame_constrt}

\subsubsection*{Five-component Formalism}

Since in our construction, a  vector boson ($V$) obtains mass by mixing with a scalar ($\varphi$) to form a single physical object, it's natural and convenient to define a unified five-component object as $V^M = (V^\mu, \varphi)$, with $M=4$ corresponding to the Goldstone component (\cite{Chen:2022gxv}).  The amplitudes for vector bosons, stripping off energy-momentum conservation, are then written as 
\be
\mathcal M(p_1,p_2,...) = \epsilon^{M_1}_{s_1}(p_1)\epsilon^{M_2}_{s_2}(p_2)... \mathcal M_{M_1 M_2...}(p_1,p_2,...), 
\ee
in which $\epsilon^{M_i}_{s_i}(p_i)$ ($i=1,2,...$) is polarization vector with spin $s_i$ and momentum $p_i$. 

Subsequently,  polarization vectors can also written in five-component form: $\epsilon^M_\sigma = (\epsilon^\mu_\sigma, \epsilon^\varphi_\sigma)$.  We can also define the five-component momentum $k^M$: $k^M=(k^\mu, \pm im_V)$.  We introduce  five-component ``matrix" $g_{MN}={\rm diag}(g_{\mu\nu},-1)$, with the  last $-1$ coming from the positive norm of the scalar $\varphi$.   This convention ensures on-shell condition ($k^2=m_V^2$) and the transverse condition ($k^\mu \epsilon_{\mu \sigma} = \pm i \epsilon^\varphi_\sigma$) follows from  the five-component form.  To eliminate the $\frac{k^\mu}{m_V}$ term, the polarization vectors and five-component momentum are chosen as
\be\label{eq:pol_vec}
\text{initial:} \  \epsilon^M_{\pm}&=&(\epsilon^{\mu}_{\pm}, 0 )^T  
\ \ \ \text{final:} \     \epsilon^{*M}_{\pm}=(\epsilon^{*\mu}_{\pm}, 0 )^T    \n \\
\text{initial:} \  \epsilon_L^M &=&(\epsilon^{\mu}_n, +i)^T 
\ \ \ \text{final:} \      \epsilon_L^{*M} =(\epsilon^{\mu}_n, -i)^T  \\
\text{initial:} \ k^M&=&(k^\mu, -im_V)  
\ \ \  \text{final:} \ k^{*M} = (k^\mu, +im_V)\n
\ee
with $\epsilon^\mu_n(k)= -\frac{m}{n\cdot k}n^\mu$, $n$ being the null vector that imposes light-cone gauge on $\epsilon^\mu$:
\begin{equation}
	n\cdot \epsilon(k,r)=0 \ \ \ \text{with} \ n^2=0
\end{equation}
If we choose $n^\mu =(1, -\frac{\vec k}{|\vec k|})$, the corresponding polarization vectors are precisely the standard ones in textbooks. 


On-shell condition, transverse condition  and orthogonal conditions can be written in five-component form as
\begin{equation}\label{eq:on_shell_cond}
	k^{*M}k_M=0	\ \ \ \ k^{*M} \epsilon_M^{\sigma} =0 \ \ \ \  \epsilon^{*M}_\sigma\epsilon_M^{\sigma'}  = -\delta_{\sigma\sigma'}
\end{equation}


On-shell gauge symmetry can also be written in five-component form as 
\begin{equation}\label{eq:MWI_2}
	k^M \mathcal M_M =0
\end{equation} 

Finally, we discuss how crossing symmetry acts on polarization vectors in  five-component form. 
It is known S-matrix is invariant under crossing symmetry, which takes the initial/final wave function with momentum $k$ and helicity $s$  to the final/initial wave function with momentum $-k$ and helicity $-s$, i.e. $u_s(k)/\bar u_s(k) \rightarrow \bar v_{-s}(-k)/v_{-s}(-k)$.   Acting on the five-component polarization vectors, a transverse polarization vector simply takes the Hermitian conjugate: $  \epsilon^{M}_\mp(k) \rightarrow \epsilon^{*M}_\mp(-k) =  \epsilon^{*M}_\mp(k) $, since they don't depend on momentum.  The situation is different for the longitudinal, $\epsilon^M_L(k) \rightarrow \epsilon^{*M}_L(-k)$. From Eq.(\ref{eq:pol_vec}), we have  $\epsilon^\mu(-k) = - \epsilon^\mu(k) $, which means that $\epsilon^{*M}_L(-k)= (-\epsilon^\mu(k), -\epsilon^{\varphi}) = - \epsilon^{M}_L(k)$.  Equivalently, $\epsilon^{*M}_L(k) = -\epsilon^{*M}_L(-k)$. Thus,   the complex conjugate for longitudinal  is equivalent to crossing symmetry up to an unphysical, overall minus sign.  This result can be useful in doing concrete, analytic calculations in five-component formalism.

\subsubsection*{Approach to Constructing Amplitudes }

Having set up the consistent conditions, we are now ready to construct massive amplitudes with them and other fundamental principles.  Here we give a sketch of the method, conveying the basic idea of our approach.  In this paper we focus on constructing four-point amplitudes, which can give us enough information to understand the underlying theory. 
Our method can be divided into  two steps: first we  construct three-point amplitudes from their massless limit by making use of the consistent conditions;  then we  construct four-point  amplitudes from three-point amplitudes by consistent factorization. 

First we analyze three-point amplitudes.  Taking the massless limit of a massive vector amplitude, its transverse and longitudinal parts   decouple from each other and  form Lorentz invariant blocks separately. Taking an amplitude with one vector boson as an example, 
\begin{equation}
	{\mathcal M}_s(V)= \epsilon^M_{\pm,L}{\mathcal M}_M(p,m_V) \overset{m_V\rightarrow 0}{\longrightarrow}
	\biggl\{
	\begin{array}{c}
		\epsilon^{\mu}_s{\mathcal M}_{\mu}(p) \ \ \ s=\pm \\
		{\mathcal M}^{\varphi}(p) \ \ \  s=L
	\end{array}
\end{equation}

Massless amplitudes can be constructed from Lorentz symmetry.  Moreover, the massless vector amplitude can be extended naturally back to massive case to include the $\epsilon_n$ part (0 in massless limit) of the longitudinal amplitude according to Eq.(\ref{eq:pol_vec}). Now we have all the elements ready to construct the massive amplitude. Based on above analysis, we propose that a massive vector amplitude can be written   as a linear combination of  massless amplitudes:
\be\label{eq:amp_decom}
\mathcal M^V_s(p,m_V) = a\  \epsilon_s^\mu \mathcal M^V_\mu(p) + b\    \mathcal M^\varphi(p)
\ee
$a,b$ are coefficients to be fixed by the   two consistent conditions.

It should be noted, however,  that there is  a potential problem when applying on-shell gauge symmetry on three-point amplitudes.  Generally it's impossible to make all  3 particles to be on-shell simultaneously, except for the special case of one particle mass larger than the sum of the two other masses.  In other words,   one of them has to be off-shell (in general kinematics).  This seems to imply on-shell gauge symmetry cannot apply on three-point amplitudes. 

However, after careful consideration, we show O.G.S is still applicable to a three-point amplitude when one particle is off-shell.   As will be clear in subsequent sections, when applying on-shell gauge symmetry by replacing $\epsilon^M$ with $k^M$,  we first make use of energy-momentum conservation to substitute $k$ with other momenta in the amplitude, then we apply on-shell condition, transverse condition and orthogonal conditions (Eq.(\ref{eq:on_shell_cond})) for the other particles.  In other words, we don't make use of any information that relevant to the on-shellness of the particle we apply O.G.S on.  Therefore,  on-shell gauge symmetry applies to an amplitude with one particle being off-shell, if  O.G.S is being applied on this off-shell particle.  This result also eliminates a motivation for complexifying the momenta.  Since there is no need,  we will keep the momenta real without complexification throughout the paper.


After  constructing three-point amplitudes, we then proceed to construct four-point amplitudes. According to consistent factorization,
when an internal line is put on-shell, amplitudes of an n-point amplitude  factorizes into amplitudes of fewer number of external points:
\be\label{eq:amp_fact}
{\mathcal M}_4 \overset{k_a^2\rightarrow m_a^2}{\longrightarrow} \sum_{s,a} {\mathcal M}_{3,a}^{s}\frac{i}{k_a^2-m_a^2}{\mathcal M}_{3,a}^{-s}
\ee
$s$ is spin index.  $a$ labels the intermediate state of which the full four-point amplitude requires summing over all possible number.  

Now we would like to reverse the factorization process and obtain the full amplitude from lower-point ones. A common way is to complexify the momenta and BCFW recursion relation.  However, as already stated above, we don't complexify momenta.  Instead we look for a natural off-shell continuation for the vector boson propagator.  This ``off-shell continuation "  can be found by writing  the summation over polarization vectors\cite{Chen:2022gxv} as
\be\label{eq:prop}
\sum_{s=\pm,0}\epsilon^M_s\epsilon^{*N}_s = -g^{MN}+\frac{k^Mn^N+n^Mk^{*N}}{n\cdot k}
\ee
The rhs of the above equation has the same form when both on-shell and off-shell, thus can be naturally used for off-shell continuation.

Next notice that $k^M$ terms give $0$ to amplitudes by on-shell gauge symmetry,  even when off-shell. Thus we can drop the $k^M$ terms for the vector propagator. The  off-shell continuation for the propagator of massive vector boson simplifies to:
\be\label{eq:pol_prop}
\frac{i\sum_{s=\pm,0}\epsilon^M_s\epsilon^{*N}_s}{k_i^2-m_i^2}\overset{k_i^2\neq m_i^2}{\longrightarrow } \frac{-ig^{MN}}{k^2_i-m_i^2}
\ee
To sum up, we can construct  amplitudes from lower points by off-shell continuation  in the following way 
\be\label{eq:amp_fact}
\sum_{s,a} {\mathcal M}_{3,a}^{s}\frac{i}{k_a^2-m_a^2}{\mathcal M}_{3,a}^{-s}
\overset{\text{off-shell}}{\longrightarrow} && {\mathcal M}_4 = \sum_a {\mathcal M}_{3,a}^M\frac{-ig_{MN}}{k_a^2-m_a^2}{\mathcal M}_{3,a}^N
\ee 

Finally, the coefficients in constructing four-point amplitudes should also be fixed by the two consistent conditions.

\section{Three-point Amplitudes}
\label{sec:three-point}

In this section we set to construct all possible three-point amplitude involving massive vector boson.  We follow the method outlined in Sec.(\ref{sec:frame_constrt}): construct massless amplitudes first, then construct massive amplitudes from massless amplitudes by the two consistent conditions. 

\subsection{ Massless three-point Amplitudes}

Let's first construct massless three-point amplitudes as basis for massive amplitudes.  Lorentz invariance implies that a scattering amplitude should be written as 
\be
\mathcal M(1,2,...,n) = \epsilon^{\mu_1}_1 \epsilon^{\mu_2}_2...\ \epsilon^{\mu_n}_n \mathcal M(1,2...,n)_{\mu_1\mu_2...\mu_n}
\ee
Polarization vectors $\epsilon_i^\mu$ undertake little group transformations, which include $U(1)$ scaling under rotation around the reference momentum, as well as the momentum shift $\epsilon^\mu \rightarrow \epsilon^\mu + \xi k^\mu$\cite{PhysRev.135.B1049} under the non-compact members of little group transformation. The latter gives massless Ward identity:
\be\label{eq:gauge_sym_massless}
k_1^{\mu_1} \epsilon^{\mu_2}_2...\epsilon^{\mu_n}_n \mathcal M(1,2...,n)_{\mu_1\mu_2...\mu_n}=0.
\ee

$\mathcal M(1,2...,n)_{\mu_1\mu_2...\mu_n}$ is a function of momenta $k_i^{\mu_i}$ and their  contraction with each other through tensor $g^{\mu\nu}$. We neglect $\epsilon^{\mu\nu\rho\sigma}$ by assuming CP symmetry is conserved.  In addition, in this paper we fix  the  energy dimension of the three-point amplitudes is  to 1, corresponding  to renormalizable Lagrangian terms only.  Finally, we use $S$ to denote scalar, $V$
to denote vector boson.

\

\

\noindent {\bf Massless $SSV$}

To maintain the dimension to be 1,  the amplitude of massless $S_1 S_2 V_3$ must be of the following form
\[{\mathcal M}_0(S_1 S_2 V_3)=(a_1k_1+a_2k_2)\cdot \epsilon_3\]
$a_1$ and $a_2$ are dimensionless parameters to be fixed.  On-shell gauge symmetry Eq.(\ref{eq:gauge_sym_massless})  constrains $a_1$ and $a_2$ to be 
\be
a_1=-a_2
\ee
Thus the amplitude for massless $1^02^03^T$ is 
\be\label{eq:3_point_massless_SSV}
{\mathcal M}_0(S_1 S_2 V_3) &=&  (k_1-k_2)\cdot \epsilon_{3} 
\ee
up to an overall factor.

Furthermore,  boson symmetry for the amplitude requires the two scalars not to be identical. Instead, the two must have different quantum numbers, which we can label as $a,b$, so that the amplitude could be written as 
\be
{\mathcal M}_0(S_1^a S_2^b V_3) = T_{ab}{\mathcal M}_0(S_1 S_2 V_3) 
\ee
with $T_{ab}=-T_{ba}$. For example, the scalars can be in a $SO(N)$ group, $T$ then is one of its generators.  There is also a special case of $N=2$, giving $SO(2)\cong U(1)$. 
In this case, the two scalars are to be identified as the antiparticles of each other.

\

\

\noindent {\bf ``Massless" $SVV$}

The amplitude of $S V V$ can be generally written as 
\be\label{eq:3_point_massless_SVV}
{\mathcal M}_0( S_1 V_2 V_3) &=& m  \epsilon_2\cdot \epsilon_3
\ee
The coefficient  $m$ has to be of  dimension $1$.  Notice there isn't anything constraining it to be a constant yet.  Generally it should be  a function of momenta: $m=m(p_1,p_2)$.
However,  we notice  Eq.(\ref{eq:3_point_massless_SVV}) doesn't satisfy massless on-shell gauge symmetry (Eq.(\ref{eq:gauge_sym_massless})), which implies  Eq.(\ref{eq:3_point_massless_SVV}) is not really a massless amplitude.  In fact, as we will see in Sec.(\ref{sec:three-point-massive}),  Eq.(\ref{eq:3_point_massless_SVV}) can be only understood as part of the massive $SVV$ amplitude that satisfies massive on-shell gauge symmetry.  This also means the amplitude should disappear in the massless limit.   Thus following strong massive-massless continuation, we must have $m\rightarrow 0$ as either one of the masses $m_i\rightarrow 0$:
\begin{equation}
	\lim_{m_i\rightarrow 0}m = 0
\end{equation}
This indicates the dependence of $m$ on particle masses in some form, which further points to common origin of $m$ and particle masses.

\

\

\noindent {\bf Massless $VVV$}

Similar to the derivation of $SSV$,  the massless amplitude of $VVV$ can be written generally as 
\[{\mathcal M}_0(V_1 V_2 V_3)=(a_{13}k_1+a_{23}k_2)\cdot \epsilon_3 \  \epsilon_1\cdot \epsilon_2  + \text{cyclic}\]
$a_{13}, a_{23}$ and etc. are dimensionless parameters to be fixed by on-shell gauge symmetry, which gives
\be
a_{13}=-a_{23}=a_{32}=-a_{12}=a_{21}=-a_{31}
\ee
$\mathcal M_0(V_1 V_2 V_3)$ becomes
\be\label{eq:3_point_massless_VVV}
{\mathcal M}_0(V_1 V_2 V_3) &=&   (\epsilon_{1}\cdot  \epsilon_{2})\ (k_1-k_2)\cdot \epsilon_{3} +\text{cyclic}
\ee
up to an overall factor. 

Moreover, boson symmetry requires a totally anti-symmetric factor for the three vector bosons. Labeling them as $a,b,c$ respectively, we arrive at the final form of  $M_0(V_1 V_2 V_3)$:
\be
{\mathcal M}_0(V_1^a V_2^b V_3^c) = f^{abc} {\mathcal M}_0(V_1 V_2 V_3)= f^{abc}\left[(\epsilon_{1}\cdot  \epsilon_{2})\ (k_1-k_2)\cdot \epsilon_{3} +\text{cyclic}\right]
\ee
$f^{abc}$ is totally anti-symmetric over $a, b, c$.

\subsection{Massive three-point  Amplitudes}
\label{sec:three-point-massive}

Having finished constructing the massless amplitudes, we now turn to the massive  amplitudes.  Apart from the trivial $SSS$ amplitude, the possible amplitudes include $SSV$, $SVV$ and $VVV$ respectively.  %
\ 

\

\noindent{\bf Massive $SSV$}

Let's start with the $SSV$ amplitude, denoted as  ${\mathcal M}(S_1S_2V_3)$. 
In the massless limit,  ${\mathcal M}(S_1S_2V_3)$ decomposes into ${\mathcal M}_0(S_1 S_2 V_3)$ and ${\mathcal M}_0(S_1 S_2 \varphi_3)$ that're  Lorentz covariant separately. 
${\mathcal M}_0(S_1 S_2 V_3)$ is given by Eq.(\ref{eq:3_point_massless_SSV}).  ${\mathcal M}(S_1 S_2 \varphi_3)$ is the amplitude for 3 scalars, which is  trivial.  

Up to an overall scaling,   the general form of  ${\mathcal M}(S_1S_2V_3) $ is 
\be\label{eq:SSV_amp}
{\mathcal M}(S_1S_2V_3)  = (k_1-k_2)\cdot \epsilon_3 - \lambda_{SS\varphi} \epsilon_3^4
\ee
$\lambda_{SS\varphi} $ is a parameter that can be fixed by massive on-shell gauge symmetry:.
\be\label{eq:gauge_symmetry_hhV}
{\mathcal M}(S_1S_2V_3)|_{\epsilon_3^M\rightarrow k_3^M} =0
\ee
which gives 
\be\label{eq:SSV_coup}
\lambda_{S_1S_2\varphi} = -i \frac{(m_1^2-m_2^2)}{m_V} 
\ee

Thus we found the solution. However, the relations between masses can still constrained by strong massive-massless continuation, which  requires that $\lambda_{SS\varphi}$  stays finite as one of the mass goes to $0$: 
\begin{equation}
	\lim_{m_i\rightarrow 0}\lambda_{SS\varphi}(m_1, m_2, m_3) \neq \infty 
\end{equation}
From it we have for $m_1$ and $m_2$:
\be\label{eq:SSV_mass}
m_1^2=c^{(0)}+ c_1^{(1)} m_V+\mathcal O(m_V^2) \ \ \ \ \ \ m_2^2=c^{(0)}+c_2^{(1)} m_V+\mathcal O(m_V^2)
\ee
This is the most general result we can obtain from the two consistent conditions, indicating at least parts of particles masses have the same physical origin.

Now let's discuss the physical meaning of the solution(Eq.(\ref{eq:SSV_coup}) and Eq.(\ref{eq:SSV_mass})) in different limits. 

First of all, if $m_V=0$, vector boson is massless. The $SS\varphi$ amplitude decouples from the gauge part.  Applying Ward identity(massless OGS) gives $m_1=m_2$. So the two scalars must have equal masses, which further implies a  $SO(2)\cong U(1)$ symmetry between $S_1$ and $S_2$.  In this case, the $SSV$ amplitude simply corresponds to QED with a massive complex scalar field. 

Second, if we impose $m_1=m_2$ while keeping $m_V\neq 0$, we obtain $\lambda_{SS\varphi} = 0$. The equal mass condition is equivalent to  imposing a $SO(2)\cong U(1)$ symmetry on $(S_1, S_2)$.  In this case the Goldstone mode is still a component of the vector boson, but never contribute to any amplitudes. In other words,  the ``Goldstone" component doesn't partake in the interactions. That means we should treat  the ``Goldstone" component as unphysical that comes from an auxiliary field. This scenario corresponds to   Stueckelberg theory. 

Third, noticing $\lambda_{SS\varphi}$ has dimension of 1, it's natural to assume it's related to the masses of the particles.  Under this assumption we can demand $\lambda_{SS\varphi}\rightarrow 0$ as $m_V\rightarrow 0$,  $m_1^2-m_2^2$ satisfies
\begin{equation}
	m_1^2-m_2^2 =  c^{(2)} m_V^2+\mathcal O(m_V^3)
\end{equation}
The scalar masses (after subtracting the constant term) are proportional to the vector mass at the lowest order, but not limited to it.

Finally, the results here only concern three-point amplitudes. In order to confirm the physical analysis above, we need to complete four-point amplitudes.

\

\

\noindent { \bf Massive: $SVV$}

Next we go on to construct the  scalar-vector-vector amplitude, denoted as ${\mathcal M}(S_3V_1V_2)$. 
In massless limit, it decomposes into ${\mathcal M}_0(S_3V_1V_2)$, ${\mathcal M}_0(S_3\varphi_1V_2)$, ${\mathcal M}_0(S_3V_1\varphi_2)$ and ${\mathcal M}_0(S_3\varphi_1\varphi_2)$. The first one is given by Eq.~(\ref{eq:3_point_massless_SVV}), the second and the third are given by Eq.~(\ref{eq:3_point_massless_SSV}), the final one is trivially a three scalar amplitude.

Thus we conclude ${\mathcal M}(S_3V_1V_2)$  has the following general form 
\be\label{eq:VVS_amp_0}
{\mathcal M}(S_3V_1V_2) =
m(\epsilon_1\cdot \epsilon_2) - a_1 (k_1-k_3)\cdot \epsilon_2 \  \epsilon_1^4 -  a_2 (k_2-k_3)\cdot \epsilon_1 \  \epsilon_2^4 + a_3  \epsilon_1^4\epsilon_2^4
\ee
$m, a_1, a_2,a_3$ are coefficients that can be determined by on-shell gauge symmetry in terms of particle masses up to an overall scaling.  

To simplify the calculation, we first apply on-shell gauge symmetry on $V_1$ by choosing $\epsilon_1^M\rightarrow k_1^M$ and setting $s_2=T$, this fixes $a_1$ to be $a_1=-\frac{im}{2m_1}$; then exchange the label of particle $1$ and particle $2$ and repeat the last step, we then get $a_2=-\frac{im}{2m_2}$; finally, we still apply on-shell gauge symmetry on $V_1$ and set $s_2=L$,  then $a_3$ is also fixed: $a_3=\frac{m m_3^2}{2m_1m_2}$. 
To summarize,  $a_1, a_2,a_3$ are fixed by on-shell gauge symmetry to be 
\be\label{eq:VVS_coup_1}
a_1=-\frac{im}{2m_1} \ \ \ \ \ a_2=-\frac{im}{2m_2} \ \ \ \ \ a_3=\frac{m m_3^2}{2m_1m_2}
\ee
We can further extract an overall factor $g_{VVS}\equiv \frac{m}{m_1}$, after which the amplitude of $SVV$ becomes
\be\label{eq:VVS_amp}
{\mathcal M}(S_3V_1V_2) =
m_1(\epsilon_1\cdot \epsilon_2) - a_1 (k_1-k_3)\cdot \epsilon_2 \  \epsilon_1^4 -  a_2 (k_2-k_3)\cdot \epsilon_1 \  \epsilon_2^4 + a_3  \epsilon_1^4\epsilon_2^4
\ee
with
\begin{equation}\label{eq:VVS_coup}
	a_1=-\frac{i}{2} \ \ \ \ \ a_2=-\frac{im_1}{2m_2} \ \ \ \ \ a_3=\frac{m_3^2}{2m_2} 
\end{equation}

Thus the massive amplitudes are uniquely fixed given the particle masses.  
Similar to  $SSV$, massive-massless continuation gives constraints on the mass parameters. 
\begin{equation}
	m=c_{1m}m_1 + \mathcal O(m_1^2) \ \ \ \ m_2= c_{12} m_1 +\mathcal O(m_1^2)  \ \ \  m_3 = c_{31} m_1 +\mathcal O(m_1^2) 
\end{equation}

\ 

\ 

\noindent { \bf Massive: $VVV$}

Next we turn to  the amplitude of three spin-1 particles. We denote it as ${\mathcal M}(V_1V_2V_3)$.  In the massless limit, the amplitude ${\mathcal M}(V_1V_2V_3)$ decomposes into 
\begin{eqnarray}
	{\mathcal M}(V_1V_2V_3)  &\longrightarrow & 
	\left \{
	\begin{array}{c}
		{\mathcal M}_0(V_1^TV_2^TV_3^T)  \\
		{\mathcal M}_0(V_1^0V_2^TV_3^T) + \text{cyclic} \\
		{\mathcal M}_0(V_1^0V_2^0V_3^T) + \text{cyclic}  \\
		{\mathcal M}_0(V_1^0V_2^0V_3^0)
	\end{array}
	\right.
\end{eqnarray}
As in the case of $SSV$ and $SVV$, we can obtain the general form of massive amplitude ${\mathcal M}(V_1V_2V_3)$ by  making continuation from the  massless limit:
\be
{\mathcal M}(V_1V_2V_3) &=&   [  (\epsilon_{1}\cdot \epsilon_{2} + a_{12}\epsilon^4_1\epsilon^4_2)\ (k_1-k_2)\cdot \epsilon_{3}+\text{cyclic}] \\ \n
&+& [b_{12}m_3\epsilon_1\cdot \epsilon_2 \epsilon_3^4 + \text{cyclic}]+ c_{123}\epsilon_1^4\epsilon_2^4\epsilon_3^4
\ee
$c_{123}$ has dimension 1, $a_{12},a_{23},a_{31}$ and $b_{12},b_{23},b_{31}$ are dimensionless coefficients. $m_1, m_2,m_3$ are taken to by physical masses of the corresponding particles.  

Again we will apply on-shell gauge symmetry to  fix the parameters.  We apply it on $k_3$ by the replacement: 
$\epsilon_3^M\rightarrow k_3^M$
and evaluate the amplitude for different helicity combinations of $k_1$ and $k_2$.

First, choosing  $s_1=T$ and $s_2=T$ gives  
\be
b_{12} &= & -i\frac{m_1^2-m_2^2}{m_3^2}  \  \  \ \text{if }\ \  m_3\neq 0 \n \\
b_{12} &=& 0   \  \  \ \text{and} \ \  m_1=m_2 \ \ \ \ \  \text{if }\ \   m_3=0
\ee

Next, choosing $s_1=T$, $s_2=L$ gives  
\be
a_{23}&=& -\frac{m_2^2+m_3^2-m_1^2}{2m_2m_3}  \  \  \ \text{if }\ \  m_2\neq 0  \ \ \text{and} \   m_3\neq 0  \n \\
a_{23} &=&  0 \ \ \  \text{if} \   \   m_3= 0 
\ee

Finally, choosing $s_1=L$, $s_2=L$ fixes $c_{123}$ to be  
\be
c_{123} = 0 
\ee
We have fixed $b_{12}, a_{23},c_{124}$. Because of cyclic invariance, we can simply rotate the indices on mass parameters and fix all the remaining parameters, which are summarized as following:

\noindent

If $m_1\neq 0,m_2\neq 0,m_3 \neq 0$:
\be\label{eq:VVV_coup_1}
&& b_{12} = -i\frac{m_1^2-m_2^2}{m_3^2}  \ \ \  \ \  \ \ \ \ \  b_{23} = -i\frac{m_2^2-m_3^2}{m_1^2} \ \ \  \ \ \ \ \ \ \  b_{31} = -i\frac{m_3^2-m_1^2}{m_2^2}  \\
&& a_{12} = -\frac{m_1^2+m_2^2-m_3^2}{2m_1m_2}  \ \ \  a_{23} = -\frac{m_2^2+m_3^2-m_1^2}{2m_2m_3} \  \ \  a_{31} = -\frac{m_3^2+m_1^2-m_2^2}{2m_3m_1}  \n  \\
&&c_{123}=0 \n
\ee

If $m_1 =0,m_2\neq 0,m_3 \neq 0$:  
\be\label{eq:VVV_coup_2}
&&b_{12}=i \ \ \ \  \  b_{23}=0 \ \ \ \  \ \    b_{31}=-i  \n  \\
&&a_{12}= 0  \  \ \ \  a_{23}=-1  \ \ \   a_{31}= 0 \\
&&c_{123}=0 \n \ \ \ \\
&& m_2=m_3 \nonumber
\ee
Notice  the additional condition of  $m_2=m_3$  has not been derived in the literature before. It seems to imply a remnant symmetry between particle $2$ and $3$ in the case of particle $1 $ being massless.

If $m_1=0,m_2= 0,m_3\neq 0$, then there is no non-trivial solution. This is commonly known as Yang's theorem.
Now we have derived the couplings the  $VVV$ amplitude with three massive particles (Eq.(\ref{eq:VVV_coup_1})),  two massive particles (Eq.(\ref{eq:VVV_coup_2})) and one massive particle (no solution) respectively.    

We can further obtain   relations between the mass parameters by applying the strong  massive-massless continuation on the couplings. This method is only useful in the case of all three being massive, in which the couplings depend on masses.
Taking $m_1/m_2/m_3$ to $0$ and requiring the couplings in Eq.(\ref{eq:VVV_coup_1}) to be finite, we deduce that   $m_1\propto m_2 \propto m_3$ at the leading order.  We can define them as functions of a common variable $v$ and have
\begin{equation}
	m_1 = c_1 v + \mathcal O(v^2) \ \ \ \   m_2=c_2 v+\mathcal O(v^2) \ \ \ \  m_3=c_3v +\mathcal O(v^3)
\end{equation}
Again, this is a signal that all particle masses have the same origin.

Before ending the  $VVV$ amplitude, we discuss two special cases of masses being equal. 

First, let's take $m_1=m_2=m_3$, we then have $b_{12}=b_{23}=b_{31}=0$ and $a_{12}=a_{23}=a_{31}=-\frac{1}{2}$. In this case the amplitude  has $SO(3)$ symmetry for the three particles.

Second, we take two of the particles to have equal masses. Let's take $m_2=m_3\neq m_1$, we have
\be\label{eq:VVV_WWZ_1}
b_{12}=i\left (1-\frac{m_1^2}{m_3^2}\right )\ \ \ \ \ \ \   \  \ \  \ \ \ \ \ \ \ \ b_{23} = 0 \ \ \ \ \  \ \ &&  b_{31}=-b_{12} \n \\
a_{12}=-\frac{m_1}{2m_3}  \ \ \ \  \ \ \ \  a_{23} =-\left (1-\frac{m_1^2}{m_3^2}\right )\ \ \ \  \ \ && a_{31}=-a_{12}
\ee
In addition we have $m_2=m_3$ automatically for  $m_1=0$ in Eq.(\ref{eq:VVV_coup_2}). 

For  $m_1>m_3$, the above two solutions correspond to $WWZ$ and $WWA$ vertices in the SM respectively.  This can be shown by defining the weak angle as  $\cos\theta_W\equiv \frac{m_3}{m_1}$ and Eq.(\ref{eq:VVV_WWZ_1})  as 
\be
b_{12}=i\frac{\sin^2\theta_W}{\cos^2\theta_W} \ \ \ \ \ \ \   \  \ \  \ \ \ \ \ \ \ \ b_{23} = 0 \ \ \ \ \  \ \ &&  b_{31}=-i\frac{\sin^2\theta_W}{\cos^2\theta_W} \n \\
a_{12}=-\frac{1}{2\cos\theta_W}  \ \ \ \  \ \ \ \  a_{23} = -\frac{\cos2\theta_W}{2\cos\theta_W}\ \ \ \  \ \ && a_{31}=-\frac{1}{2\cos\theta_W}  
\ee
This is indeed the couplings of $WWZ$, including $\varphi\varphi V$ and $\varphi V V$ vertices.

\newpage

\section{Massive Four-point Amplitudes}
\label{sec:four-point}

As discussed before, three-point amplitudes are not generally viable by kinematics.  The minimal number for n-point amplitudes to be kinematically allowed generically is  $n=4$.  Thus  in order to have consistent full theories,  it is essential to construct up to four-point amplitudes at least, which is what we are doing in this section.  
The approach of constructing four-point amplitudes is explained in Sec.(\ref{sec:frame_constrt}), we mainly make use of consistent factorization   in addition to the two conditions of on-shell gauge symmetry and massive-massless continuation. 

According to the principle of quantum mechanics, to construct a four-point amplitude,  we need to sum over all possible three-point amplitudes with different intermediate states and four-point contact terms. Moreover, bosonic symmetry (for the massless limit) constrains the number of types of particles (with different quantum numbers) for three-point amplitudes. For $SSV$,  the two scalars are required to be of different types; for $VVV$, the three vector bosons also must be of different types.  
As a result, we can then classify different theories according to the number of vector bosons $n_V$ and the number of scalars $n_S$, which fix the possible three-point amplitudes contributing to the final four-point amplitude: 
\begin{itemize}
	\item $n_V <  3$
	\begin{itemize}
		\item $n_V=1, n_S\geq 2$: $SSV$ 
		\item $n_V\leq 2$, $n_S=1$: $VVS$ 
		\item $n_V\leq 2$, $n_S\geq 2$: $VVS$,  $SSV$
	\end{itemize}
	\item $n_V\geq 3$
	\begin{itemize}
		\item  $n_V \geq 3, n_S=0$: $VVV$
		\item  $n_V \geq 3, n_S=1$: $VVV$, $VVS$
		\item  $n_V \geq 3, n_S\geq 2 $: $VVV$, $VVS$ and $SSV$
	\end{itemize}
\end{itemize}
For every case with at least one scalar,  we can also add $SSS$.   

For  $n_V<  2$,  our strategy is to first construct the amplitudes from $VVS$, then add $SSV$. In this way we will be able to cover all three cases.  For $n_V\geq 3$,  we use  a different strategy. We first try to construct four-point amplitudes from $VVV$ alone; then expand to incorporate $VVS$.   We will also  neglect $SSV$ and leave it to future work. 

We will go into the massive case directly. Massless four-point amplitudes can be automatically included in the massive case, by taking masses to $0$ and keeping the relevant degrees of freedom only.  So they will not be discussed separately.

\subsection{$n_V\leq 2$: Four-point Amplitudes from  $VVS$ and $SSV$}

Here we construct four-point amplitudes from  $VVS$,  $SSV$ and $SSS$.  Except for the  trivial $SSSS$, there are two possible four-point amplitudes  $VVVV$ and $SSVV$, which we will construct  one by one.   After that, we will classify and discuss the underlying physics of the solutions.

\subsubsection {$VVVV$  from  $VVS$}       

To construct the amplitude of  $VVVV$ we need only $VVS$.  To simplify the analysis without losing generality, we focus on studying the case of   $n_V=1, n_S=1$,   so that there is only one independent $VVS$.   Then we will generalize to two scalars in the end. 

The total amplitude equals the sum of s,t,u channels and the contact term:
\be
\mathcal M_{\text{tot}}= \mathcal M_s + \mathcal M_t+ \mathcal M_u+\mathcal M_c
\ee
$\mathcal M_s$, $\mathcal M_t$ and $\mathcal M_t$  are $s,t,u$ channels from three-point vertices. By consistent factorization we can obtain their forms when the intermediate states approach on-shell. Plugging in the three-point amplitudes for $VVS$(Eq.(\ref{eq:VVS_amp} and \ref{eq:VVS_coup})) and $SSV$(Eq.(\ref{eq:SSV_amp} and \ref{eq:SSV_coup})), we get
\be\label{eq:VVVV_fac}
\mathcal M_s(V_1 V_2 V_3 V_4) &\overset{p_{12}^2\rightarrow m_S^2}{\longrightarrow}  & \mathcal M(V_1V_2S_{12})\frac{1}{p_{12}^2-m_S^2} \mathcal M(-S_{12}V_3V_4) \n \\
\mathcal M_t(V_1 V_2 V_3 V_4) &\overset{p_{13}^2\rightarrow m_V^2}{\longrightarrow}  & \sum_{s_{13}=\pm,0}\mathcal  M(V_1V_3S_{13})\frac{1}{p_{13}^2-m_V^2}\mathcal M(-S_{13}V_2V_4) \\
\mathcal M_u(V_1 V_2 V_3V_4) &\overset{p_{14}^2\rightarrow m_V^2}{\longrightarrow}  & \sum_{s_{14}=\pm,0}\mathcal M(V_1V_4S_{14})\frac{1}{p_{14}^2-m_V^2}\mathcal M(-S_{14}V_2V_3) \n 
\ee
Here $\mathcal M(VVS)$  should be understood to be multiplied by an overall coupling $g_{VVS}$.  $m_S$ and $m_V$ are the masses of the scalar and the vector boson respectively. The two vector bosons in  $VVS$ are identical, therefore have the same mass.  
The off-shell continuation for scalar is trivial, so amplitudes for s, t, u channels  are simply
\be\label{eq:VVSS-1}
\mathcal M_s(V_1 V_2 V_3 V_4) &=& \mathcal M(V_1V_2S_{12})\frac{1}{p_{12}^2-m_S^2} \mathcal  M(-S_{12}V_3V_4) \n \\
\mathcal M_t(V_1 V_2 V_3 V_4) &= & \mathcal M(V_1V_3S_{13})\frac{1}{p_{13}^2-m_V^2}\mathcal M(-S_{13}V_2V_4) \\
\mathcal M_u(V_1 V_2 S_3S_4) &=  &\mathcal M(V_1V_4S_{14})\frac{1}{p_{14}^2-m_V^2}\mathcal M(-S_{14}V_2V_3) \n 
\ee

To have the full amplitude we also need the contact term  $\mathcal  M_c$, which  can be generally written as 
\be
\mathcal M_c&=&a_1 \epsilon_1\cdot \epsilon_2 \ \epsilon_3\cdot \epsilon_4 + a_2 \epsilon_1\cdot \epsilon_3 \ \epsilon_2\cdot \epsilon_4 + a_3 \epsilon_1\cdot\epsilon_4 \ \epsilon_3\cdot \epsilon_2 \n \\
&&+ b_{11} \epsilon_1\cdot \epsilon_2 \ \epsilon_3^4\cdot \epsilon_4^4+ b_{12} \epsilon_1^4\cdot \epsilon^4_2 \ \epsilon_3\cdot \epsilon_4\n \\
&&+ b_{21} \epsilon_1\cdot \epsilon_3 \ \epsilon_2^4\cdot \epsilon_4^4+ b_{22} \epsilon_1^4\cdot \epsilon^4_3 \ \epsilon_2\cdot \epsilon_4  \\
&&+ b_{31} \epsilon_1\cdot \epsilon_4 \ \epsilon_2^4\cdot \epsilon_3^4+ b_{32} \epsilon_1^4\cdot \epsilon^4_4 \ \epsilon_2\cdot \epsilon_3 \n \\
&&+ c\  \epsilon_1^4\epsilon_2^4 \epsilon_3^4 \epsilon_4^4 \n
\ee 
$a_1, a_2,a_3$, $b_{11}, b_{12}, b_{21},b_{22}, b_{31},b_{32},c$ are dimensionless coefficients that are to be fixed by on-shell gauge symmetry for the total amplitude. $c$ is simply the self-coupling of Goldstone bosons, which we also  label as $\lambda_{\varphi^4}\equiv c$.  So up to an overall coupling $g_{VVS}^2$,  there are 10 free parameter in total.  

To fix the coefficients we apply on-shell gauge symmetry on $\epsilon_1^M$.  For $s,t,u$ channels we get
\be\label{eq:VVS_VVVV_ons}
\mathcal M_s^{\epsilon_1^M\rightarrow k_1^M }&=& \frac{i}{2}g^2_{VVS}\epsilon_2^4\left(m_V\epsilon_3\cdot \epsilon_4 +\frac{i}{2}(2p_3+p_4)\cdot \epsilon_4\ \epsilon_3^4+\frac{i}{2}(2p_4+p_2)\cdot \epsilon_3\ \epsilon_4^4+\frac{m_S^2}{2m_V}\epsilon_3^4\epsilon_4^4\right)\n \\
\mathcal M_t^{\epsilon_1^M\rightarrow k_1^M }&=& \frac{i}{2} g_{VVS}^2\epsilon_3^4\left(m_V\epsilon_2\cdot \epsilon_4 +\frac{i}{2}(2p_2+p_4)\cdot \epsilon_4\ \epsilon_2^4+\frac{i}{2}(2p_4+p_3)\cdot \epsilon_2\ \epsilon_4^4+\frac{m_S^2}{2m_V}\epsilon_2^4\epsilon_4^4\right) \\
\mathcal M_u^{\epsilon_1^M\rightarrow k_1^M }&=&\frac{i}{2} g_{VVS}^2 \epsilon_4^4\left(m_V\epsilon_2\cdot \epsilon_3 +\frac{i}{2}(2p_2+p_3)\cdot \epsilon_3\ \epsilon_2^4+\frac{i}{2}(2p_3+p_4)\cdot \epsilon_2\ \epsilon_3^4+\frac{m_S^2}{2m_V}\epsilon_2^4\epsilon_3^4\right) \n 
\ee
For the contact channel we get 
\begin{eqnarray}
	\mathcal M_c^{\epsilon_1^M\rightarrow k_1^M} &=&  a_1 k_1\cdot \epsilon_2\ \epsilon_3\cdot \epsilon_4 + a_2 \epsilon_1\cdot \epsilon_3 \ \epsilon_2\cdot \epsilon_4 + a_3 \epsilon_1\cdot\epsilon_4 \ \epsilon_3\cdot \epsilon_2 \n \\
	&&+ b_{11} \epsilon_1\cdot \epsilon_2 \ \epsilon_3^4\cdot \epsilon_4^4+ b_{12} \epsilon_1^4\cdot \epsilon^4_2 \ \epsilon_3\cdot \epsilon_4\n \\
	&&+ b_{21} \epsilon_1\cdot \epsilon_3 \ \epsilon_2^4\cdot \epsilon_4^4+ b_{22} \epsilon_1^4\cdot \epsilon^4_3 \ \epsilon_2\cdot \epsilon_4  \\
	&&+ b_{31} \epsilon_1\cdot \epsilon_4 \ \epsilon_2^4\cdot \epsilon_3^4+ b_{32} \epsilon_1^4\cdot \epsilon^4_4 \ \epsilon_2\cdot \epsilon_3 \n \\
	&&+ c\  \epsilon_1^4\epsilon_2^4 \epsilon_3^4 \epsilon_4^4 \n
\end{eqnarray}

Adding them up and we obtain the following results:
\be\label{eq:VVVV_VVS}
a_1 &=& a_2=a_3=0 \n \\
g_{\varphi\varphi VV}= b_{11}&=&b_{12}=b_{21}=b_{22}=b_{31}=b_{31}=\frac{1}{2}g_{VVS}^2 \\
c &\equiv&\lambda_{\varphi^4}=\frac{3}{4}g_{VVS}^2\frac{m_S^2}{m_V^2}\n
\ee
From Eq.(\ref{eq:VVVV_VVS}), $g_{\varphi\varphi VV}$ (i.e.$b_{ij}$) and $\lambda_{\varphi^4}$ are related by $\lambda_{\varphi^4}=\frac{3}{2}\frac{m^2_S}{m_V^2}g_{\varphi\varphi VV}$. Thus all four-point couplings are fixed by $g_{VVS}$, an overall coupling. 

If we have more than one scalar, the results can be generalized naturally. For example,  let's assume we have two scalars $S^a$ and $S^b$ that couple with $V$,  the solution of on-shell gauge symmetry becomes 
\begin{eqnarray}
	a_1 &=& a_2=a_3=0 \n \\
	b_{11}&=&b_{12}=b_{21}=b_{22}=b_{31}=b_{31}=\frac{1}{2}\left (g_{VVS^a}^2+g_{VVS^b}^2\right ) \\
	c &\equiv&\lambda_{\varphi^4}=\frac{3}{4m_V^2}\left(g_{VVS^a}^2m_a^2+g^2_{VVS^b}m_b^2\right)\n
\end{eqnarray}
In order to completely fix the couplings, we need two parameters as input, which we can  choose as $g_{VVS^a}$ and the ratio of $r_1\equiv \frac{g_{VVS^a}}{g_{VVV^b}}$.

\subsubsection{ $SSVV$ from $SSV$ and $VVS$}

Here we construct $SSVV$, which  is from either $VVS$   or $SSV$. 
$SSV$ also already implies  the existence of $VVS$ and $SSS$.  Our strategy is to first construct $SSVV$ with $SSV$, $VVS$ and $SSS$, then take $g_{VVS}=0$ and $g_{SSV} =0 $ respectively to see the minimal structure for the amplitude.

Again, to avoid unnecessary complication, we focus on the case with minimal number of particles, i.e. with only one type of vector boson and  two types of scalars. This  implies one $SSV$ amplitude  $\mathcal M(S^aS^bV)$; as well as two $VVS$ amplitudes of $\mathcal M(VVS^a)$ and $ \mathcal M(VVS^b)$.  We focus on analyzing the example of $\mathcal M(S^a_aS^a_2V_3V_4)$, as there is no essential difference in other four-point amplitudes. 

The total amplitude of $M(S^a_1S^a_2V_3V_4)$ can be constructed from s, t, u channels and the contact term:
\be
\mathcal M_{\text{tot}}= \mathcal M_s+\mathcal M_t+\mathcal M_u+\mathcal M_c
\ee
with

\be 
\mathcal  M_s(S^a_1S^a_2V_3V_4) &=& \sum_{i=a,b}\mathcal M(S_1^aS_2^aS^i_{12})\frac{1}{p_{12}^2-m_i^2}\mathcal M(S^i_{12}V_3V_4) \n \\
\mathcal M_t(S^a_1S^a_2V_3V_4)  &=& \mathcal M(S_1^aV_3S^b_{13})\frac{1}{p_{13}^2-m_b^2}\mathcal M(S^b_{13}S_2^aV_4) \\
&&+\mathcal M(S_1^aV_3V_{13})^{M_{13}}\frac{-g_{M_{13}N_{13}}}{p_{13}^2-m_V^2}\mathcal M^{N_{13}}(V_{13}S_2^aV_4^{s_4})  \n\\ 
\mathcal M_u(S^a_1S^a_2V_3V_4)  &=& \mathcal M(S_1^aV_4S^b_{14})\frac{1}{p_{14}^2-m_i^2}\mathcal M(S^b_{14}S_2^aV_3)\n \\
&&+\mathcal M(S_1^aV_4V_{14})^{M_{14}}\frac{-g_{M_{14}N_{14}}}{p_{14}^2-m_V^2}\mathcal M^{N_{14}}(V_{14}S_2^aV_3^{s_3})  \n
\ee
Notice we have to sum over all possible particles in the intermediate particles for every channel. 

In addition, the contact term gives:
\be
\mathcal M_c(S^a_1S^a_2V_3V_4)=g_{S^aS^aVV}\ \epsilon_3\cdot \epsilon_4 +\lambda_{SS\varphi^a\varphi^a}\  \epsilon_3^4\epsilon_4^4
\ee
Thus the full amplitude of $S^a_1S^a_2V_3V_4$ has 7 free parameters: $\lambda_{S^aS^aS^a}, \lambda_{S^aS^aS^b}$,  $g_{VVS^a}$, $g_{VVS^b}$, $g_{S^aS^bV}$, $g_{S^aS^aVV}$ and $\lambda_{SS\varphi^a\varphi^a}$. Those parameters are subject to additional constraints from on-shell gauge symmetry. Replacing $\epsilon_3^M$ with $p_3^M$, we have:
\be
\mathcal M_{\text{tot}}^{\epsilon_3^M\rightarrow p_3^M}=0
\ee
which gives the following solution:
\be\label{eq:SSVV_sol}
g_{S^aS^aVV}&=& -2\left(g_{S^aS^bV}^2-\frac{g_{VVS^a}^2}{4}\right)\n \\
\frac{1}{2}(\lambda_{S^aS^aS^a}g_{VVS^a}+\lambda_{S^aS^aS^b}g_{VVS^b})-\lambda_{S^aS^a\varphi\varphi}m_V&=&
g^2_{S^aS^bV}\frac{m_a^2-m_b^2}{m_V}-\frac{m_a^2}{2m_V}g_{VVS^a}^2
\ee
The first line comes from taking $s_4=T$,  the second line comes from taking $s_4=L$.  Exchanging $a$ and $b$, we further get 
\be\label{eq:SSVV_sol_2}
g_{S^bS^bVV}&=& -2\left(g_{S^aS^bV}^2-\frac{g_{VVS^b}^2}{4}\right)\n \\
\frac{1}{2}(\lambda_{S^aS^aS^a}g_{VVS^a}+\lambda_{S^aS^aS^b}g_{VVS^b})-\lambda_{S^aS^a\varphi\varphi}m_V&=&
g^2_{S^aS^bV}\frac{m_a^2-m_b^2}{m_V}-\frac{m_a^2}{2m_V}g_{VVS^a}^2
\ee

Now let's analyze the solution in Eq.(\ref{eq:SSVV_sol}) by taking various limits. 

First, if we reduce the number of scalars to one, which means $g_{S^aS^bV}=0$ and $\lambda_{S^aS^aS^b}=0$. Setting $S^a=S$, Eq.(\ref{eq:SSVV_sol}) becomes
\be\label{eq:VVSS_sol_noSb}
g_{SSVV}&=&\frac{1}{2}g_{VVS}^2\n\\
\lambda_{\varphi\varphi SS} &=& g_{VVS}\frac{\lambda_{SSS}}{m_V}+\frac{m_S^2}{2m_V^2}g_{VVS}^2
\ee 
In this case, the three-point amplitudes that contribute to the full  amplitude are $VVS$ and $SSS$.   There is only one free parameter $\lambda_{SSS}$ after setting the overall coupling $g_{VVS}=1$.  Considering that $\lambda_{SSSS}$ also has no constraints on it, all couplings can be fixed by particle masses except for the scalar self-couplings ($\lambda_{SSS}$ and $\lambda_{SSSS}$).  However, strong massive-massless continuation on Eq.(\ref{eq:VVSS_sol_noSb}) (second line)  constrains the behavior of $\lambda_{SSS}$ at the limit of $m_V\rightarrow 0$:
\begin{equation}
	\lim_{m_V\rightarrow 0} \lambda_{SSS}=0
\end{equation}
So $\lambda_{SSS}$ doesn't exist in the massless limit.  Also it indicates that $\lambda_{SSS}$ has the same physical origin as the particle masses. 

Next, we take  $g_{VVS^a}=g_{VVS^b}= 0$ and $m_a=m_b$.  Eq.(\ref{eq:SSVV_sol}) then reduces to 
\be\label{eq:SSVV_sol_noVVS}
g_{S^aS^aVV}=-2g_{S^aS^bV}^2 \ \ \ \ \   \lambda_{S^aS^a\varphi\varphi}&=& 0 
\ee
In this case,  $\varphi$ doesn't contribute  to the amplitude. Physically it means that    we should treat it as an auxiliary mode, instead of a physical, dynamic degree of freedom.  Moreover, scalar couplings $\lambda_{S^aS^aS^a/S^b}$ decouple from the amplitude.    

Finally, we analyze  the general case in Eq.(\ref{eq:SSVV_sol}).   The 3 conditions (Eq.(\ref{eq:SSVV_sol}) and  the freedom of overall scaling)  reduce the number of free parameters  from 7 to 4, only enough to eliminate  the two four-point couplings and one three-point coupling.  In addition we  treat $\lambda_{S^aS^aS^a}$ and $\lambda_{S^aS^aS^b}$ as input. We then have 5 relevant parameters constrained by 3 conditions.    In summary,  we still need $5-3=2$ free parameters.  To fix the remaining two parameters, we can set the ratios:
\begin{equation}\label{eq:ratio}
	r_1\equiv \frac{g_{VVS^a}}{g_{VVS^b}} \ \ \ \  \ 
	r_2\equiv \frac{g_{VVS^a}}{g_{S^aS^bV}}
\end{equation}
as input, in addition to $\lambda_{S^aS^aS^a}$ and $\lambda_{S^aS^aS^b}$.

\subsubsection{Underlying  Theories  and Models }

Now having finished constructing four-point amplitudes for $n_V <  3$,  we then set to discuss the underlying physical theories and models for those constructions. 

From one vector boson and one scalar, we can construct  only one three-point amplitude of $VVS$ and three four-point amplitudes of $SSSS$, $SSVV$ and $VVVV$.    In constructing those amplitudes,  we found all couplings are fixed by the masses, except for the scalar couplings of $SSS$ and $SSSS$.  Moreover, by applying strong massive-massless continuation on both three-point and four-point amplitudes,  we found that when one of the masses goes to $0$, other particles' masses and  $g_{VVS}$ also go to $0$.  This implies a common origin for the all particle masses and couplings of mass dimension, a mark of SSB.  However, the relations between the masses are generally nonlinear.  Physically, the nonlinearity between masses and  scalar couplings being free parameters  indicate  the Higgs potential is not limited to the terms of dim $\leq$ 4.   In order to fix dim $\ge$ 4  terms for Higgs potential, we need to go beyond four-point in constructing amplitudes, thus is not within the scope of this paper.  For example, we can add dim-6 SMEFT operators, or consider Coleman-Weinberg potential.  In conclusion, the underlying constructible theory with one massive vector boson and one scalar, 
is massive QED with SSB, with the corresponding Lagrangian being composed of a vector field  and a complex scalar field.  Moreover,  the specific shape of the Higgs potential cannot be determined.

From one vector boson and two scalars, we can construct three-point amplitudes of $SVV$ and $SSV$, as well as four-point amplitudes of $SSSS$, $SSVV$ and $VVVV$.   In constructing those amplitudes,  we need two more parameters in Eq.(\ref{eq:ratio}) as input, apart from masses,  to completely fix the couplings except for scalar self-couplings.   The other aspects are the same as the case of one vector boson and one scalar.  Thus the underlying theory is still massive gauge theory with SSB, but with more than one Higgs field.   For a specific model, we can take  the $U(1)$ version of two-Higgs-doublet model as an example.  The two parameters in Eq.(\ref{eq:ratio}) fix the angle between the two vevs(vacuum expectation values) ($\beta$) and the mixing angle of the two neutral and CP-even scalars ($\alpha$).   This case can also be a basis for analyzing more general cases, such as the non-abelian cases with mixing between Higgs fields and etc.

Finally, there is a special limit for one vector boson and two scalars, in which any couplings involving the Goldstone $\varphi$  are zero.   In the cases we studied, there are two necessary conditions. The first is  that the two scalars have equal masses. The second is all  $VVS$ couplings are $0$.  The underlying theory of this scenario is Stueckelberg theory, in which the vector boson has mass without  SSB. $\varphi$ is only a pure gauge that doesn't contribute to any amplitude.

\subsection{$n_V\geq 3$: Four-point Amplitudes from $VVV$, $VVS$ and $SSV$ }

For $n_V\geq 3$, we have possible three-point amplitudes of $VVV$, $VVS$ and $SSV$. The possible four-point amplitudes constructed are $VVVV$, $VVSS$ and $SSSS$. $SSSS$  is not constrained by on-shell gauge symmetry, so we will skip it.  
Similarly, we also won't construct  $VVSS$, which  has been constructed in the case of $n_V=1$.  There is no new physical insights for $n_V\geq 3$.     So in summary we will  only construct $VVVV$ with $VVV$ and $VVS$.  
Our strategy is to construct $VVVV$ from  the three-point amplitude $VVV$ alone first (by assuming there is no scalar), to see if it's possible to be compatible with consistent conditions.  If it doesn't work,  we will then add $VVS$. 

\subsubsection{$VVVV$ from  $VVV$}

Here we set to construct $VVVV$ from $VVV$ and $VVS$.  We start with  $n_S=0$, meaning no $VVS$ first. We also put the ``color" explicitly  for the amplitudes.  For VVV we have  $\mathcal M(V_1^aV_2^bV_3^c)=f^{abc}\mathcal M(V_1V_2V_3)$, with $f^{abc}$ to be antisymmetric tensor.  For simplicity, we set all vector bosons
to have the same mass.

The condition of consistent factorization, combined by a natural off-shell continuation, ensures the total four-point amplitude $\mathcal M(V^aV^bV^cV^d)$ equals the sum of s, t, u channel and the contact channel:
\be
\mathcal M_{\text{tot}}(V^a_1V^b_2V^c_3V^d_4) =\mathcal M_s+\mathcal M_t+\mathcal M_u+\mathcal M_c
\ee
with $\mathcal M_s, \mathcal M_t, \mathcal M_u$ being 
\be 
\mathcal M_s(V^a_1V^b_2V^c_3V^d_4) &=& \sum_{e}\mathcal M(V_1^aV_2^bV^e_{12})^{M_{12}} \frac{-g_{M_{12}N_{12}}}{p_{12}^2-m_V^2}\mathcal M^{N_{12}}(V^e_{12}V^c_3V^d_4) \n \\
\mathcal M_t(V^a_1V^b_2V^c_3V^d_4)  &=&\sum_e \mathcal M(V_1^aV^c_3V_{13}^e)^{M_{13}}\frac{-g_{M_{13}N_{13}}}{p_{13}^2-m_V^2}\mathcal M^{N_{13}}(V^e_{13}V^b_2V^d_4)  \\ 
\mathcal M_u(V^a_1V^b_2V^c_3V^d_4)  &=& \sum_e \mathcal M(V_1^aV^d_4V_{14}^e)^{M_{14}}\frac{-g_{M_{14}N_{14}}}{p_{14}^2-m_V^2}\mathcal M^{N_{14}}(V^e_{14}V^b_2V^c_3) 
\ee
The contact term $\mathcal M_c$ has no pole, therefore should be the linear combination of all possible terms from  Lorentz contraction between $\epsilon_i^\mu,\epsilon_i^4$ with $i=1,2,3,4$. 
\be\label{eq:VVV_mc}
\mathcal M_c &=&  a_{12,34} \ \epsilon_1\cdot \epsilon_2 \ \epsilon_3\cdot \epsilon_4
+a_{13,24}\ \epsilon_1\cdot \epsilon_3 \ \epsilon_2\cdot \epsilon_4 
+a_{14,23}\ \epsilon_1\cdot \epsilon_4 \ \epsilon_2\cdot \epsilon_3 \n \\
&&+ b_{12,34}\ \epsilon_1\cdot\epsilon_2 \epsilon_3^4\epsilon_4^4
+ b_{13,24}\ \epsilon_1\cdot \epsilon_3 \ \epsilon_2^4\epsilon_4 ^4
+b_{14,23}\ \epsilon_1\cdot \epsilon_4 \ \epsilon_2^4\epsilon_3^4 \\
&&+ b_{34,12}\ \epsilon_3\cdot \epsilon_4  \epsilon_1^4\epsilon_2^4 
+ b_{24,13}\ \epsilon_2\cdot\epsilon_4 \ \epsilon_1^4 \epsilon_3^4 
+b_{23,14}\ \epsilon_2\cdot \epsilon_3 \ \epsilon_1^4\epsilon_4^4 \n \\
&&+ \lambda_{\varphi^4}  \epsilon^4_1\epsilon_2^4\epsilon_3^4\epsilon_4^4 \n
\ee
The coefficients are to be determined by the additional constraint of on-shell gauge symmetry. 

We first apply on-shell gauge symmetry on $\mathcal M(V_1V_2V_{12})$ with one of the other legs, say  $p_{12}=p_1+p_2$, being off-shell. Choosing the particle for replacing $\epsilon^M$ with $p^M$ to be $p_1$, we then have:
\be
\mathcal M_{\epsilon^M_1\rightarrow p_1^M}^{p_{12}^2\neq m_V^2} (V_1V_2V_{12})= \left(\epsilon_2\cdot \epsilon_3-\frac{1}{2}\epsilon_2^4\epsilon_3^4\right)(p_{12}^2-m_V^2) + p_{12}^M\cdot \epsilon^{12*}_{M}(-\frac{i}{2}m_V\epsilon_2^4-p_3\cdot \epsilon_2)
\ee
Making use of it, we can obtain the solution for $\mathcal M_{\text{tot}}(V^a_1V^b_2V^c_3V^d_4) _{\epsilon_1^M\rightarrow p_1^M}=0$. However, because the solution is a little complicated. We simplify it by solving on-shell gauge symmetry with $s_2,s_3,s_4$ being transverse(T) and longitudinal(L) respectively. 

First, taking $s_2=T, s_3=T,s_4=T$. In this case only terms such as $\epsilon_1\cdot \epsilon_2 \ \epsilon_3\cdot \epsilon_4$ contribute to $\mathcal M_c$, which can be written as 
\be\label{eq:VVVV_mc1}
&& \mathcal M_c^{\epsilon_1\rightarrow p_1}(s_2=T, s_3=T,s_4=T)\n \\
&=& a_{12,34} \ p_1\cdot \epsilon_2 \ \epsilon_3\cdot \epsilon_4
+a_{13,24}\ p_1\cdot \epsilon_3 \ \epsilon_2\cdot \epsilon_4 
+a_{14,23}\ p_1\cdot \epsilon_4 \ \epsilon_2\cdot \epsilon_3
\ee

We evaluate $(\mathcal M_s +\mathcal M_t+\mathcal M_u)_{\epsilon_1^M\rightarrow p_1^M}$ first, which gives 
\be
&&(\mathcal M_s+\mathcal M_t+\mathcal M_u)_{\epsilon_1^M\rightarrow p_1^M}  \n \\
& =& -(f^{abe}f^{cde} + f^{ace}f^{dbe}+f^{ade}f^{bce})
\left[\epsilon_2\cdot \epsilon_3\ (p_1+2p_2)\cdot \epsilon_4 -2\epsilon_2\cdot \epsilon_4\  p_2\cdot p_3 +2\epsilon_3\cdot \epsilon_4 \ (-p_1-p_4)\cdot \epsilon_2\right] \n \\
&&+ f^{abe}f^{cde}\left[p_1\cdot \epsilon_3\ \epsilon_2\cdot \epsilon_4-p_1\cdot \epsilon_4\ \epsilon_2\cdot \epsilon_3\right]\n\\
&&+f^{ace}f^{dbe}\left[p_1\cdot \epsilon_4\ \epsilon_2\cdot \epsilon_3-p_1\cdot \epsilon_2\ \epsilon_3\cdot \epsilon_4 \right]\\
&&+f^{ade}f^{bce}\left[p_1\cdot \epsilon_2\ \epsilon_3\cdot \epsilon_4-p_1\cdot \epsilon_3\ \epsilon_2\cdot \epsilon_4 \right]\n
\ee
Comparing with Eq.(\ref{eq:VVVV_mc1}), we find that there is no solution for on-shell gauge symmetry,  unless there is an additional condition:
\be\label{eq:jacobi}
f^{abe}f^{cde} + f^{ace}f^{dbe}+f^{ade}f^{bce}=0
\ee
in which case we have the solution for coefficients:
\be
a_{12,34} = f^{ace}f^{bde}-f^{ade}f^{bce}; \ \  a_{13,24} = f^{ade}f^{bce}-f^{abe}f^{cde};  \ \ a_{14,23} = f^{abe}f^{cde}-f^{ace}f^{dbe}
\ee
we then obtain $\mathcal M_c$ with all particles being transverse:
\be
\mathcal M_c
&=&-f^{abe}f^{cde}\left[\epsilon_1\cdot \epsilon_3\ \epsilon_2\cdot \epsilon_4-\epsilon_1\cdot \epsilon_4\ \epsilon_2\cdot \epsilon_3\right]\n\\
&&-f^{ace}f^{dbe}\left[\epsilon_1\cdot \epsilon_4\ \epsilon_2\cdot \epsilon_3-\epsilon_1\cdot \epsilon_2\ \epsilon_3\cdot \epsilon_4 \right]\\
&&-f^{ade}f^{bce}\left[\epsilon_1\cdot \epsilon_2\ \epsilon_3\cdot \epsilon_4-\epsilon_1\cdot \epsilon_3\ \epsilon_2\cdot \epsilon_4 \right]\n
\ee

Eq.(\ref{eq:jacobi}) is simply Jacobi identity, indicating that the involved vector bosons with different quantum numbers  belong to the same Lie group. Notice all transverse cases reduce  to massless vector scattering directly when $m_V\rightarrow 0$, thus we arrive at the conclusion that the only consistent massless theory with vector boson scattering is Yang-Mills theory. 

Then we continue to apply on-shell gauge symmetry  with one or several particles being longitudinal.  We found, however, there is no non-trivial solution for on-shell gauge symmetry.  In particular, the solutions for $b_{ij, kl}$ by  applying O.G.S on $s_2=L$, $s_3=T, s_4=T$ and $s_2=L, s_3=L, s_4=T$ contradict with each other.   

Choosing $s_2=L$, $s_3=T$ and $s_4=T$,  on-shell gauge symmetry is satisfied if the coefficients in Eq.(\ref{eq:VVV_mc}) satisfy the following conditions:
\be
b_{34,12} = 0; \ \  b_{24,13} = 0; \  \ b_{23,14}  =0
\ee
which fixes the $VV\varphi\varphi$ vertices. By permutation symmetry we also obtain all $b_{ij,kl}$ coefficients:
\be\label{eq:VVV_wrong_sol_1}
b_{12,34}= b_{34,12} =0 ; \ \  b_{13,24} = b_{24,13}=0; \ \ b_{14,23} =b_{23,14} =0 
\ee

On the other hand, the solution of $b_{ij.kl}$ for O.G.S by choosing $s_2=L, s_3=L, s_4=T$ is 
\be\label{eq:VVV_wrong_sol_2}
b_{14,23} = - f^{abe}f^{cde}  = f^{ace}f^{dbe} \ \  \  b_{34,12}=0 \ \  \ b_{24,13}=0 
\ee
It's impossible for  Eq.(\ref{eq:VVV_wrong_sol_1}) and Eq.(\ref{eq:VVV_wrong_sol_2}) to be correct at the same time, except for the trivial solution of $f^{abe}f^{cde}  = f^{ace}f^{dbe}=0$.  Therefore, it's impossible to satisfy on-shell gauge symmetry for $\mathcal M(V^aV^bV^cV^d)$ with vector bosons only.

For the case of elementary particles, the only solution  is to add additional degrees of freedom in the theory. For $\mathcal M(V^aV^bV^cV^d)$, the only possibility is to add additional scalars, and consequently the three-point amplitude $\mathcal M(VVS)$. For simplicity, we only add one scalar to the theory.  This gives 3 additional $s,t,u$ channels contributing to $\mathcal M(V^aV^bV^cV^d)$ with the scalar to be the intermediate particle. The total amplitude now becomes
\be
\mathcal M_{\text{tot}}(V^a_1V^b_2V^c_3V^d_4) =\mathcal M_s+\mathcal M_t+\mathcal M_u+\mathcal M^S_s+\mathcal M^S_t+\mathcal M^S_u+\mathcal M_c
\ee
The scalar channels' contribution to on-shell gauge symmetry was obtained in Eq.(\ref{eq:VVS_VVVV_ons}). 

Applying on-shell gauge symmetry again for $s_2=L,s_3=T,s_4=T$, the related coefficients are modified to
\be
b_{34,12}=\frac{1}{2}g_{V^aV^bS}g_{V^cV^dS}
\ee
Using permutation, we can obtain all $b_{ij,kl}$ coefficients:
\be\label{eq:VVV_LTT}
b_{12,34}&=& b_{34,12}=\frac{1}{2}g_{V^aV^bS}g_{V^cV^dS} \nonumber \\
b_{13,24}& =& b_{24,13}=\frac{1}{2}g_{V^aV^cS}g_{V^bV^dS};  \\
b_{14,23} &=&b_{23,14} = \frac{1}{2}g_{V^aV^dS}g_{V^bV^cS} \nonumber
\ee

We then apply on-shell gauge symmetry to $s_2=L,s_3=L,s_4=T$, which gives the following constraint on $VVS$ vertices:
\be\label{eq:VVV_LLT_condition}
0&=&  -\frac{1}{4} f^{abe}f^{cde} \epsilon^4_2\epsilon^4_3 (2p_2+p_4)\cdot \epsilon_4 + \frac{1}{4} f^{ace}f^{db e} \epsilon^4_2\epsilon^4_3 (2p_3+p_4)\cdot \epsilon_4  \\
&+&b_{34,12}\epsilon_3\cdot \epsilon_4 (-im_V) \epsilon_2^4 + b_{24,13} \epsilon_3\cdot \epsilon_4 (-im_V) \epsilon_3^4 +b_{14,23} p_1\cdot \epsilon_4 \epsilon_2^4\epsilon_3^4  \nonumber \\
&+& g_{V^aV^bS}g_{V^cV^dS} [\frac{i}{2}m_V\epsilon_2^4(\epsilon_3\cdot\epsilon_4) +\frac{1}{4}(2p_3+p_4)\cdot \epsilon_4\epsilon_2^4\epsilon_3^4]
\nonumber \\
&+& g_{V^aV^cS} g_{V^dV^bS} [\frac{i}{2}m_V\epsilon_3^4 (\epsilon_2\cdot \epsilon_4)+\frac{1}{4}(p_2+p_4)\cdot \epsilon_4\epsilon_2^4\epsilon_3^4 \nonumber ]
\ee
Plugging in Eq.(\ref{eq:VVV_LTT}), we obtain the following relations between $g_{VVS}$ and $f^{abc}$:
\begin{eqnarray}\label{eq:VVV_LLT}
	f^{abe}f^{cde} = 	g_{V^aV^cS} g_{V^bV^dS} - g_{V^aV^dS}g_{V^bV^cS} \n \\
	f^{ace}f^{dbe} =  -g_{V^aV^bS}g_{V^cV^dS}+	g_{V^aV^dS} g_{V^bV^cS} 
\end{eqnarray}
$g_{V^iV^j S}$ is symmetric with $i\leftrightarrow j$ for equal vector boson masses, e.g. $g_{V^bV^aS}=g_{V^aV^bS} $.  We can also apply O.G.S on $s_2=L,s_3=T,s_4=L$ and $s_2=T,s_3=L,s_4=L$, but there is no new independent results.  Eq.(\ref{eq:VVV_LLT}) gives us the only constraints between the group structure coefficients  $f^{abc}$ and the $VVS$ couplings.

Finally, we choose $s_2=L,s_3=L,s_4=L$ and apply on-shell gauge symmetry on $p_1$, which fixes $\lambda_{\varphi^4}$ to be
\begin{equation}\label{eq:VVV_LLL}
	\lambda_{\varphi^4} = \frac{1}{4}\frac{m_S^2}{m_V^2}(g_{V^aV^bS}g_{V^cV^dS}+g_{V^aV^cS}g_{V^dV^bS}+g_{V^aV^dS}g_{V^bV^cS})
\end{equation}

So we have successfully  constructed massive $VVVV$ from $VVV$, with all vector boson masses being equal.  We proved a scalar is needed for the construction to be consistent.  We believe this conclusion holds for general cases of vector boson masses.    All couplings are constrained by particle masses and the group structure constants $f^{abc}$ in Eq.(\ref{eq:VVV_LTT}, \ref{eq:VVV_LLT} and \ref{eq:VVV_LLL}).  

Those solutions can be further simplified if we introduce a mild and reasonable assumption.  Because all vector bosons have the same mass and there is only one scalar, it is reasonable to assume all $VVS$ couplings are equal, i.e.  $g\equiv g_{V^aV^bS}=g_{V^cV^dS}=g_{V^aV^cS}=g_{V^bV^dS}=g_{V^aV^dS}=g_{V^bV^cS}$. However, in this case, Eq.(\ref{eq:VVV_LLT}) becomes $f^{abe}f^{cde}= f^{ace}f^{dbe} = f^{ade}f^{cbe}=0$, reducing the solution  to the abelian case.   Nevertheless, there is a way to save the situation by making one of the  $VVS$ coupling equal to $0$. For example, we can take 
\begin{eqnarray}\label{eq:VVV_eq_mass_gVVS}
	&&g_{V^aV^bS}=0 \ \ \text{or}\ \   g_{V^cV^dS}=0 \nonumber \\
	&&g_{VVS}\equiv g_{V^aV^cS}=g_{V^bV^dS}=g_{V^aV^dS}=g_{V^bV^cS}
\end{eqnarray} 
The condition eliminates one scalar mediated channel. 
Eq.(\ref{eq:VVV_LLT}) then reduces to 
\begin{equation}\label{eq:VVV_LLT_eq_mas}
	f^{abe}f^{cde}=0 \ \ \ \ \  g_{VVS}^2=f^{ace}f^{dbe} =  f^{ade}f^{bce}
\end{equation}
We  see that one of vector boson mediated channel is also eliminated. As a result,  the $VVVV$ amplitude has only $s$ and $t$ channels that can be organized as  planar diagrams.  This can be checked in  simplest case of  $SU(2)$ group. In the $W^{\pm}$ basis, the diagrams correspond to exactly   Eq.(\ref{eq:VVV_LLT_eq_mas}) with $f^{abc}=\epsilon^{abc}$ and $a, b, c= \pm, 3$. 
It's also interesting to point out that   Eq.(\ref{eq:VVV_LLT_eq_mas}) is also the solution for 
\begin{equation}
	g_{V^aV^bS}g_{V^cV^dS}=g_{V^aV^dS} g_{V^bV^cS} = 0 \ \ \ \ \ g_{VVS}\equiv g_{V^aV^cS}=g_{V^bV^dS}
\end{equation}
In other words, there is only one scalar mediated channel while all relevant $VVS$ couplings are equal.

\subsubsection{The Underlying Theory with Amplitudes from $VVV$}

For the number of vector bosons equal to or larger than 3, we have the three-point $VVV$, from which  we try to construct $VVVV$. For all particles being transverse,  it leads to Jacobi identity for the couplings of $VVV$, thus establishing the classic result of massless Yang-Mills theory. For other polarizations, we  found that an additional scalar is needed for self-consistency of the amplitude as required on-shell gauge symmetry.  After adding the scalar and the subsequent three-point amplitude $VVS$,  we are able to construct $VVVV$ with all four-point couplings and $VVS$ couplings fixed by masses and group structure constants (Eq.(\ref{eq:VVV_LTT}), Eq.(\ref{eq:VVV_LLT}) and Eq.(\ref{eq:VVV_LLL})).   Those results, combined with the results of $VVV$ in Sec.(\ref{sec:three-point-massive}), are enough for us to conclude  that,  for $VVVV$ with the number of vector bosons larger than or equal to 3 and all particles are elementary, the only possible underlying theory is Yang-Mills theory with SSB.  In particular, we can conclude  that Stueckelberg  theory cannot be applied on theory with the vertex of $VVV$, i.e. non-abelian gauge theory. The reason is that the Goldstone mode $\varphi$ always contributes non-trivially to amplitudes in this case. Therefore, it's impossible to  have Stueckelberg  theory,  in which $\varphi$ is simply  pure gauge.

Furthermore,  we also discuss the  solutions under the reasonable assumption of all $VVS$ couplings being equal. We found that non-trivial solutions require one of the $s, t, u$ channels of both one scalar mediated and  vector boson mediated amplitudes  cannot exist.  The  couplings are fixed as in Eq.(\ref{eq:VVV_eq_mass_gVVS}) and Eq.(\ref{eq:VVV_LLT_eq_mas}). The corresponding amplitudes then have only planar diagrams.  A specific example is  $SU(2)$ group  in the basis of $W^\pm$.

\section{Conclusion}
\label{sec:conclusion}

In this paper we set to construct massive amplitudes with elementary particles.  We focus on amplitudes with vector bosons and scalars up to four-point.   For this purpose, we propose two consistent conditions for the amplitudes:  on-shell gauge symmetry and strong massive-massless continuation(Item~(\ref{item:consis_cond})). The latter condition is to ensure the particles are elementary by imposing on  the analytic property of amplitudes as functions of masses. The former condition is simply massive Ward identity   that becomes Goldstone equivalence theorem in high energy limit. However, we argued in the paper that it can be seen as coming from Lorentz symmetry and unitairty, combined with the principle of manifest massive-massless continuation.  By mixing vector boson with an auxiliary scalar, constrained by the mixing condition in Eq.(\ref{eq: mix_cond_V-phi}),  we eliminated the state with negative norm. Both the mixing condition and the corresponding polarization vectors have smooth massless limits.  Most importantly, we naturally obtain on-shell gauge symmetry.   

After proposing the two consistent conditions,  we continue to construct three-point and four-point amplitudes.  To construct three-point, we apply the principle of massive-massless continuation on amplitudes, concluding that  massive amplitudes can be written as linear combinations of the massless counter parts(Eq.(\ref{eq:amp_decom})).  So the basic strategy is to construct the massless amplitudes first,   then apply the two consistent conditions to construct massive amplitudes, which fix the coefficients and their relations.  With this strategy, we successfully construct all possible three-point massive amplitudes: $SSV$, $SVV$ and $VVV$.  All couplings, up to an overall scaling, are fixed in terms of masses.  For $VVV$ we also discuss the solution with different mass combinations and reconstruct the $VVV$ amplitudes in the SM. Finally, when one of the masses goes to $0$, other masses and dimensional couplings also must go to 0, indicating common physical origin.

To construct four-point amplitudes, we make use of consistent factorization that the four-point amplitude factorizes into products of  three-point amplitudes   when either of $s,t,u$ channels goes on-shell for one of the particles.  Starting from four-point amplitudes when  the intermediates states are on-shell, we then extend the propagator off-shell continuously to obtain the amplitudes at general kinematics(Eq.(\ref{eq:amp_fact})).  We classify our construction according to the number of vector bosons $n_V$:  for $n_V<3$,  we construct $VVSS$ and $VVVV$.  With only one scalar we found all couplings are fixed in terms of masses, except for scalar self-couplings.  The underlying theory is massive scalar QED with SSB.   With two scalars,   we also need the ratio of $\frac{g_{VVS^a}}{g_{VVS^b}}$ and $\frac{g_{VVS^a}}{g_{S^aS^bV}}$.  Physically, the underlying theory for one vector boson and two scalars are the $U(1)$ version of two-Higgs-Doublet-Model
(2HDM).  The two ratios are the mixing angles between two VEVs and between two neutral, CP-even scalars. There is also a special case when the scalars have equal masses.  In this case the ``Goldstone" mode $\varphi$ does not contribute  to the amplitude.  This corresponds to Stueckelberg theory, in which there is no need for symmetry breaking for  a vector boson to have mass.

For $n_V\ge 3$, we construct $VVVV$ from $VVV$.  We found that in order to satisfy on-shell gauge symmetry, additional scalars must be aded.  After adding an additional scalar and   vertex $VVS$, the amplitude of $VVVV$ is successfully constructed.  All four-point couplings except scalar self-couplings are then fixed. $VVS$ couplings are further constrained by the group structure constants in Eq.(\ref{eq:VVV_LLT}). We therefore conclude  the only possible underlying theory for $VVVV$ from $VVS$ is Yang-Mills theory with SSB, when all particles are elementary.  We also discuss the solutions in some special conditions such as the case of all $VVS$ couplings being equal, the $SU(2)$ group and etc.  

In comparison with other similar works in the literature,  our approach of constructing amplitudes has a few advantages.  First of all, our method has  manifest gauge symmetry at the level of amplitudes by making Goldstone bosons part of physical spectrum.  Second, we essentially derive on-shell gauge symmetry from Lorentz symmetry and some reasonable assumptions of mixing between vector boson and scalar, thus putting our approach on a firm theoretical foundation. Third,  the  condition of strong massive-massless continuation allows us to distinguish theories with elementary particles from composite particles through  analytic properties of amplitudes. Finally,  our results are also more complete compared with other approaches, as demonstrated by the conclusions of Stueckelberg theory, particles masses have the same physical origin,  Higgs self-couplings are modifies by dim $>$ 4 terms of the Higgs potential and etc. 

Our results  open the door to many new directions, while also  leaving a few questions to be answered.   To name a few topics for future research:   four-point amplitudes with fermions remain to be constructed;  for $VVVV$ from $VVV$, we have only studied the case of equal masses.  The amplitude with  general masses are still yet to be fully constructed.    Furthermore, it's also interesting to  go beyond four-point amplitudes  to see what can be learned.

\section*{Acknowledgement}
Junmou Chen is supported by National Natural Science Foundation of China under Grant No 12205118.

\appendix

\section{Derivation of Vector-Scalar Kinematic Lagrangian}
\label{sec:mix_Lagran}

Here we derive the kinematic Lagrangian terms for $V-\varphi$ mixing, from the assumptions of both $V$ and $\varphi$ have mass $m_V$, and physical condition $\partial^\mu V_\mu = m_V \varphi$.

The kinematic Lagrangian related to the vector boson is 
\[
\mathcal L =-\frac{1}{4} (\partial_\mu V_\nu -\partial_\nu V_\mu)^2 +\frac{1}{2} m_V^2 V_\mu V^\mu + \mathcal L_{V-\varphi}
\]
$\mathcal L_{V-\varphi}$ is unknown.  Euler-Lagrangian equation gives
\[
-(\partial^2 + m_V^2) V_\mu + \partial_\mu(\partial\cdot V) + (V-\varphi \ \text{terms}) = 0
\]
Applying the physical condition $\partial^\mu V_\mu=m_V\varphi$ fixes the $V-\varphi$ term above to be $-m_V\partial_\mu \varphi$, meaning $\mathcal L_{V-\varphi}$ is 
\[
\mathcal L_{V-\varphi}= m_V V_\mu \partial^\mu \varphi
\]

In similar way we can derive the Lagrangian for $\varphi$, of which the general form is
\[
\mathcal L_\varphi = \frac{1}{2}(\partial_\mu \varphi)^2 -\frac{1}{2}m_{\varphi}^2 \varphi^2
\]
The equation of motion for $\varphi$ is then
\[
(\partial^2 + m_\varphi^2)\varphi + m_V\partial^\mu V_\mu =0
\]
Again plugging physical condition eliminates $V_\mu$ and gives 
\[
(\partial^2 + m_\varphi^2-m_V^2)\varphi =0 
\]
This fixes $m_\varphi =0$. 

To sum up, the kinematic Lagrangian for $V$ and $\varphi$ with mixing is 
`
\begin{eqnarray}
	\mathcal L_{V^2} &=& -\frac{1}{4}(\partial_\mu V_\nu -\partial_\nu V_\mu)^2 +\frac{1}{2}m_V^2V_\mu^2 \nonumber \\
	\mathcal L_{V-\varphi}  &=& m_V V_\mu \partial^\mu \varphi \\
	\mathcal L_{\varphi^2} &=&  \frac{1}{2}\partial_\mu \varphi \partial^\mu \varphi \nonumber
\end{eqnarray}

\bibliography{construct-vector-arxiv}

\begin{thebibliography}{42}
\expandafter\ifx\csname natexlab\endcsname\relax\def\natexlab#1{#1}\fi
\expandafter\ifx\csname bibnamefont\endcsname\relax
  \def\bibnamefont#1{#1}\fi
\expandafter\ifx\csname bibfnamefont\endcsname\relax
  \def\bibfnamefont#1{#1}\fi
\expandafter\ifx\csname citenamefont\endcsname\relax
  \def\citenamefont#1{#1}\fi
\expandafter\ifx\csname url\endcsname\relax
  \def\url#1{\texttt{#1}}\fi
\expandafter\ifx\csname urlprefix\endcsname\relax\def\urlprefix{URL }\fi
\providecommand{\bibinfo}[2]{#2}
\providecommand{\eprint}[2][]{\url{#2}}

\bibitem[{\citenamefont{Weinberg}(1964{\natexlab{a}})}]{PhysRev.134.B882}
\bibinfo{author}{\bibfnamefont{S.}~\bibnamefont{Weinberg}},
  \emph{\bibinfo{title}{{Feynman Rules for Any Spin. II. Massless Particles}}},
  \bibinfo{journal}{Phys. Rev.} \textbf{\bibinfo{volume}{134}},
  \bibinfo{pages}{B882} (\bibinfo{year}{1964}{\natexlab{a}}).

\bibitem[{\citenamefont{Weinberg}(1964{\natexlab{b}})}]{PhysRev.135.B1049}
\bibinfo{author}{\bibfnamefont{S.}~\bibnamefont{Weinberg}},
  \emph{\bibinfo{title}{{Photons and {G}ravitons in {$S$}-Matrix {T}heory:
  Derivation of Charge Conservation and Equality of Gravitational and Inertial
  Mass}}}, \bibinfo{journal}{Phys. Rev.} \textbf{\bibinfo{volume}{135}},
  \bibinfo{pages}{B1049} (\bibinfo{year}{1964}{\natexlab{b}}).

\bibitem[{\citenamefont{Weinberg}(1964{\natexlab{c}})}]{WEINBERG1964357}
\bibinfo{author}{\bibfnamefont{S.}~\bibnamefont{Weinberg}},
  \emph{\bibinfo{title}{{Derivation of Gauge Invariance and the Equivalence
  Principle from Lorentz Invariance of the S- matrix}}},
  \bibinfo{journal}{Physics Letters} \textbf{\bibinfo{volume}{9}},
  \bibinfo{pages}{357 } (\bibinfo{year}{1964}{\natexlab{c}}), ISSN
  \bibinfo{issn}{0031-9163}.

\bibitem[{\citenamefont{Britto et~al.}(2005)\citenamefont{Britto, Cachazo,
  Feng, and Witten}}]{Britto:2005fq}
\bibinfo{author}{\bibfnamefont{R.}~\bibnamefont{Britto}},
  \bibinfo{author}{\bibfnamefont{F.}~\bibnamefont{Cachazo}},
  \bibinfo{author}{\bibfnamefont{B.}~\bibnamefont{Feng}}, \bibnamefont{and}
  \bibinfo{author}{\bibfnamefont{E.}~\bibnamefont{Witten}},
  \emph{\bibinfo{title}{{Direct proof of tree-level recursion relation in
  Yang-Mills theory}}}, \bibinfo{journal}{Phys. Rev. Lett.}
  \textbf{\bibinfo{volume}{94}}, \bibinfo{pages}{181602}
  (\bibinfo{year}{2005}), \eprint{hep-th/0501052}.

\bibitem[{\citenamefont{Benincasa and Cachazo}(2007)}]{Benincasa:2007xk}
\bibinfo{author}{\bibfnamefont{P.}~\bibnamefont{Benincasa}} \bibnamefont{and}
  \bibinfo{author}{\bibfnamefont{F.}~\bibnamefont{Cachazo}},
  \emph{\bibinfo{title}{{Consistency Conditions on the S-Matrix of Massless
  Particles}}} (\bibinfo{year}{2007}), \eprint{0705.4305}.

\bibitem[{\citenamefont{McGady and Rodina}(2014)}]{McGady:2013sga}
\bibinfo{author}{\bibfnamefont{D.~A.} \bibnamefont{McGady}} \bibnamefont{and}
  \bibinfo{author}{\bibfnamefont{L.}~\bibnamefont{Rodina}},
  \emph{\bibinfo{title}{{Higher-spin massless $S$-matrices in
  four-dimensions}}}, \bibinfo{journal}{Phys. Rev. D}
  \textbf{\bibinfo{volume}{90}}, \bibinfo{pages}{084048}
  (\bibinfo{year}{2014}), \eprint{1311.2938}.

\bibitem[{\citenamefont{Schuster and Toro}(2009)}]{Schuster:2008nh}
\bibinfo{author}{\bibfnamefont{P.~C.} \bibnamefont{Schuster}} \bibnamefont{and}
  \bibinfo{author}{\bibfnamefont{N.}~\bibnamefont{Toro}},
  \emph{\bibinfo{title}{{Constructing the Tree-Level Yang-Mills S-Matrix Using
  Complex Factorization}}}, \bibinfo{journal}{JHEP}
  \textbf{\bibinfo{volume}{06}}, \bibinfo{pages}{079} (\bibinfo{year}{2009}),
  \eprint{0811.3207}.

\bibitem[{\citenamefont{Arkani-Hamed et~al.}(2021)\citenamefont{Arkani-Hamed,
  Huang, and Huang}}]{Arkani-Hamed:2017jhn}
\bibinfo{author}{\bibfnamefont{N.}~\bibnamefont{Arkani-Hamed}},
  \bibinfo{author}{\bibfnamefont{T.-C.} \bibnamefont{Huang}}, \bibnamefont{and}
  \bibinfo{author}{\bibfnamefont{Y.-t.} \bibnamefont{Huang}},
  \emph{\bibinfo{title}{{Scattering amplitudes for all masses and spins}}},
  \bibinfo{journal}{JHEP} \textbf{\bibinfo{volume}{11}}, \bibinfo{pages}{070}
  (\bibinfo{year}{2021}), \eprint{1709.04891}.

\bibitem[{\citenamefont{Cornwall et~al.}(1973)\citenamefont{Cornwall, Levin,
  and Tiktopoulos}}]{Cornwall:1973tb}
\bibinfo{author}{\bibfnamefont{J.~M.} \bibnamefont{Cornwall}},
  \bibinfo{author}{\bibfnamefont{D.~N.} \bibnamefont{Levin}}, \bibnamefont{and}
  \bibinfo{author}{\bibfnamefont{G.}~\bibnamefont{Tiktopoulos}},
  \emph{\bibinfo{title}{{Uniqueness of spontaneously broken gauge theories}}},
  \bibinfo{journal}{Phys. Rev. Lett.} \textbf{\bibinfo{volume}{30}},
  \bibinfo{pages}{1268} (\bibinfo{year}{1973}), \bibinfo{note}{[Erratum:
  Phys.Rev.Lett. 31, 572 (1973)]}.

\bibitem[{\citenamefont{Cornwall et~al.}(1974)\citenamefont{Cornwall, Levin,
  and Tiktopoulos}}]{Cornwall:1974km}
\bibinfo{author}{\bibfnamefont{J.~M.} \bibnamefont{Cornwall}},
  \bibinfo{author}{\bibfnamefont{D.~N.} \bibnamefont{Levin}}, \bibnamefont{and}
  \bibinfo{author}{\bibfnamefont{G.}~\bibnamefont{Tiktopoulos}},
  \emph{\bibinfo{title}{{Derivation of Gauge Invariance from High-Energy
  Unitarity Bounds on the s-Matrix}}}, \bibinfo{journal}{Phys. Rev. D}
  \textbf{\bibinfo{volume}{10}}, \bibinfo{pages}{1145} (\bibinfo{year}{1974}),
  \bibinfo{note}{[Erratum: Phys.Rev.D 11, 972 (1975)]}.

\bibitem[{\citenamefont{Llewellyn~Smith}(1973)}]{LlewellynSmith:1973yud}
\bibinfo{author}{\bibfnamefont{C.~H.} \bibnamefont{Llewellyn~Smith}},
  \emph{\bibinfo{title}{{High-Energy Behavior and Gauge Symmetry}}},
  \bibinfo{journal}{Phys. Lett. B} \textbf{\bibinfo{volume}{46}},
  \bibinfo{pages}{233} (\bibinfo{year}{1973}).

\bibitem[{\citenamefont{Lee et~al.}(1977)\citenamefont{Lee, Quigg, and
  Thacker}}]{Lee:1977yc}
\bibinfo{author}{\bibfnamefont{B.~W.} \bibnamefont{Lee}},
  \bibinfo{author}{\bibfnamefont{C.}~\bibnamefont{Quigg}}, \bibnamefont{and}
  \bibinfo{author}{\bibfnamefont{H.~B.} \bibnamefont{Thacker}},
  \emph{\bibinfo{title}{{The Strength of Weak Interactions at Very
  High-Energies and the Higgs Boson Mass}}}, \bibinfo{journal}{Phys. Rev.
  Lett.} \textbf{\bibinfo{volume}{38}}, \bibinfo{pages}{883}
  (\bibinfo{year}{1977}).

\bibitem[{\citenamefont{Conde and Marzolla}(2016)}]{Conde_2016}
\bibinfo{author}{\bibfnamefont{E.}~\bibnamefont{Conde}} \bibnamefont{and}
  \bibinfo{author}{\bibfnamefont{A.}~\bibnamefont{Marzolla}},
  \emph{\bibinfo{title}{Lorentz constraints on massive three-point
  amplitudes}}, \bibinfo{journal}{Journal of High Energy Physics}
  \textbf{\bibinfo{volume}{2016}} (\bibinfo{year}{2016}), ISSN
  \bibinfo{issn}{1029-8479},
  \urlprefix\url{http://dx.doi.org/10.1007/JHEP09(2016)041}.

\bibitem[{\citenamefont{Bachu and Yelleshpur}(2020)}]{Bachu_2020}
\bibinfo{author}{\bibfnamefont{B.}~\bibnamefont{Bachu}} \bibnamefont{and}
  \bibinfo{author}{\bibfnamefont{A.}~\bibnamefont{Yelleshpur}},
  \emph{\bibinfo{title}{On-shell electroweak sector and the higgs mechanism}},
  \bibinfo{journal}{Journal of High Energy Physics}
  \textbf{\bibinfo{volume}{2020}} (\bibinfo{year}{2020}), ISSN
  \bibinfo{issn}{1029-8479},
  \urlprefix\url{http://dx.doi.org/10.1007/JHEP08(2020)039}.

\bibitem[{\citenamefont{Choi and Jeong}(2022)}]{Choi:2021szj}
\bibinfo{author}{\bibfnamefont{S.~Y.} \bibnamefont{Choi}} \bibnamefont{and}
  \bibinfo{author}{\bibfnamefont{J.~H.} \bibnamefont{Jeong}},
  \emph{\bibinfo{title}{{Constructing the covariant three-point vertices
  systematically}}}, \bibinfo{journal}{Phys. Rev. D}
  \textbf{\bibinfo{volume}{105}}, \bibinfo{pages}{016016}
  (\bibinfo{year}{2022}), \eprint{2111.08236}.

\bibitem[{\citenamefont{Bachu}(2023)}]{Bachu:2023fjn}
\bibinfo{author}{\bibfnamefont{B.}~\bibnamefont{Bachu}},
  \emph{\bibinfo{title}{{Spontaneous Symmetry Breaking from an On-Shell
  Perspective}}} (\bibinfo{year}{2023}), \eprint{2305.02502}.

\bibitem[{\citenamefont{Liu and Yin}(2022)}]{Liu:2022alx}
\bibinfo{author}{\bibfnamefont{D.}~\bibnamefont{Liu}} \bibnamefont{and}
  \bibinfo{author}{\bibfnamefont{Z.}~\bibnamefont{Yin}},
  \emph{\bibinfo{title}{{Gauge invariance from on-shell massive amplitudes and
  tree-level unitarity}}}, \bibinfo{journal}{Phys. Rev. D}
  \textbf{\bibinfo{volume}{106}}, \bibinfo{pages}{076003}
  (\bibinfo{year}{2022}), \eprint{2204.13119}.

\bibitem[{\citenamefont{Lai et~al.}(2024)\citenamefont{Lai, Liu, and
  Terning}}]{Lai:2023upa}
\bibinfo{author}{\bibfnamefont{H.-Y.} \bibnamefont{Lai}},
  \bibinfo{author}{\bibfnamefont{D.}~\bibnamefont{Liu}}, \bibnamefont{and}
  \bibinfo{author}{\bibfnamefont{J.}~\bibnamefont{Terning}},
  \emph{\bibinfo{title}{{The constructive method for massive particles in
  QED}}}, \bibinfo{journal}{JHEP} \textbf{\bibinfo{volume}{06}},
  \bibinfo{pages}{086} (\bibinfo{year}{2024}), \eprint{2312.11621}.

\bibitem[{\citenamefont{Christensen}(2024)}]{Christensen:2024xzs}
\bibinfo{author}{\bibfnamefont{N.}~\bibnamefont{Christensen}},
  \emph{\bibinfo{title}{{Perturbative unitarity and the four-point vertices in
  the constructive standard model}}}, \bibinfo{journal}{Phys. Rev. D}
  \textbf{\bibinfo{volume}{109}}, \bibinfo{pages}{116014}
  (\bibinfo{year}{2024}), \eprint{2403.07977}.

\bibitem[{\citenamefont{Dong et~al.}(2023)\citenamefont{Dong, Ma, Shu, and
  Zhou}}]{Dong_2023}
\bibinfo{author}{\bibfnamefont{Z.-Y.} \bibnamefont{Dong}},
  \bibinfo{author}{\bibfnamefont{T.}~\bibnamefont{Ma}},
  \bibinfo{author}{\bibfnamefont{J.}~\bibnamefont{Shu}}, \bibnamefont{and}
  \bibinfo{author}{\bibfnamefont{Z.-Z.} \bibnamefont{Zhou}},
  \emph{\bibinfo{title}{The new formulation of higgs effective field theory}},
  \bibinfo{journal}{Journal of High Energy Physics}
  \textbf{\bibinfo{volume}{2023}} (\bibinfo{year}{2023}), ISSN
  \bibinfo{issn}{1029-8479},
  \urlprefix\url{http://dx.doi.org/10.1007/JHEP09(2023)101}.

\bibitem[{\citenamefont{Liu et~al.}(2023)\citenamefont{Liu, Ma, Shadmi, and
  Waterbury}}]{Liu_2023}
\bibinfo{author}{\bibfnamefont{H.}~\bibnamefont{Liu}},
  \bibinfo{author}{\bibfnamefont{T.}~\bibnamefont{Ma}},
  \bibinfo{author}{\bibfnamefont{Y.}~\bibnamefont{Shadmi}}, \bibnamefont{and}
  \bibinfo{author}{\bibfnamefont{M.}~\bibnamefont{Waterbury}},
  \emph{\bibinfo{title}{An eft hunter’s guide to two-to-two scattering: Heft
  and smeft on-shell amplitudes}}, \bibinfo{journal}{Journal of High Energy
  Physics} \textbf{\bibinfo{volume}{2023}} (\bibinfo{year}{2023}), ISSN
  \bibinfo{issn}{1029-8479},
  \urlprefix\url{http://dx.doi.org/10.1007/JHEP05(2023)241}.

\bibitem[{\citenamefont{Bresciani et~al.}(2025)\citenamefont{Bresciani, Levati,
  and Paradisi}}]{Bresciani:2025toe}
\bibinfo{author}{\bibfnamefont{L.~C.} \bibnamefont{Bresciani}},
  \bibinfo{author}{\bibfnamefont{G.}~\bibnamefont{Levati}}, \bibnamefont{and}
  \bibinfo{author}{\bibfnamefont{P.}~\bibnamefont{Paradisi}},
  \emph{\bibinfo{title}{{Amplitudes and partial wave unitarity bounds}}}
  (\bibinfo{year}{2025}), \eprint{2504.12855}.

\bibitem[{\citenamefont{Balkin et~al.}(2022)\citenamefont{Balkin, Durieux,
  Kitahara, Shadmi, and Weiss}}]{Balkin_2022}
\bibinfo{author}{\bibfnamefont{R.}~\bibnamefont{Balkin}},
  \bibinfo{author}{\bibfnamefont{G.}~\bibnamefont{Durieux}},
  \bibinfo{author}{\bibfnamefont{T.}~\bibnamefont{Kitahara}},
  \bibinfo{author}{\bibfnamefont{Y.}~\bibnamefont{Shadmi}}, \bibnamefont{and}
  \bibinfo{author}{\bibfnamefont{Y.}~\bibnamefont{Weiss}},
  \emph{\bibinfo{title}{On-shell higgsing for efts}}, \bibinfo{journal}{Journal
  of High Energy Physics} \textbf{\bibinfo{volume}{2022}}
  (\bibinfo{year}{2022}), ISSN \bibinfo{issn}{1029-8479},
  \urlprefix\url{http://dx.doi.org/10.1007/JHEP03(2022)129}.

\bibitem[{\citenamefont{Durieux et~al.}(2020)\citenamefont{Durieux, Kitahara,
  Machado, Shadmi, and Weiss}}]{Durieux_2020}
\bibinfo{author}{\bibfnamefont{G.}~\bibnamefont{Durieux}},
  \bibinfo{author}{\bibfnamefont{T.}~\bibnamefont{Kitahara}},
  \bibinfo{author}{\bibfnamefont{C.~S.} \bibnamefont{Machado}},
  \bibinfo{author}{\bibfnamefont{Y.}~\bibnamefont{Shadmi}}, \bibnamefont{and}
  \bibinfo{author}{\bibfnamefont{Y.}~\bibnamefont{Weiss}},
  \emph{\bibinfo{title}{Constructing massive on-shell contact terms}},
  \bibinfo{journal}{Journal of High Energy Physics}
  \textbf{\bibinfo{volume}{2020}} (\bibinfo{year}{2020}), ISSN
  \bibinfo{issn}{1029-8479},
  \urlprefix\url{http://dx.doi.org/10.1007/JHEP12(2020)175}.

\bibitem[{\citenamefont{Balasubramanian
  et~al.}(2023)\citenamefont{Balasubramanian, Chakraborty, Rudra, and
  Saha}}]{Balasubramanian_2023}
\bibinfo{author}{\bibfnamefont{M.~K.~N.} \bibnamefont{Balasubramanian}},
  \bibinfo{author}{\bibfnamefont{K.}~\bibnamefont{Chakraborty}},
  \bibinfo{author}{\bibfnamefont{A.}~\bibnamefont{Rudra}}, \bibnamefont{and}
  \bibinfo{author}{\bibfnamefont{A.~P.} \bibnamefont{Saha}},
  \emph{\bibinfo{title}{On-shell supersymmetry and higher-spin amplitudes}},
  \bibinfo{journal}{Journal of High Energy Physics}
  \textbf{\bibinfo{volume}{2023}} (\bibinfo{year}{2023}), ISSN
  \bibinfo{issn}{1029-8479},
  \urlprefix\url{http://dx.doi.org/10.1007/JHEP06(2023)037}.

\bibitem[{\citenamefont{Wu and Zhu}(2022)}]{Wu_2022}
\bibinfo{author}{\bibfnamefont{C.}~\bibnamefont{Wu}} \bibnamefont{and}
  \bibinfo{author}{\bibfnamefont{S.-H.} \bibnamefont{Zhu}},
  \emph{\bibinfo{title}{Massive on-shell recursion relations for n-point
  amplitudes}}, \bibinfo{journal}{Journal of High Energy Physics}
  \textbf{\bibinfo{volume}{2022}} (\bibinfo{year}{2022}), ISSN
  \bibinfo{issn}{1029-8479},
  \urlprefix\url{http://dx.doi.org/10.1007/JHEP06(2022)117}.

\bibitem[{\citenamefont{Ni et~al.}(2025)\citenamefont{Ni, Wang, Wu, and
  Yu}}]{Ni:2025xkg}
\bibinfo{author}{\bibfnamefont{Y.-H.} \bibnamefont{Ni}},
  \bibinfo{author}{\bibfnamefont{Y.-N.} \bibnamefont{Wang}},
  \bibinfo{author}{\bibfnamefont{C.}~\bibnamefont{Wu}}, \bibnamefont{and}
  \bibinfo{author}{\bibfnamefont{J.-H.} \bibnamefont{Yu}},
  \emph{\bibinfo{title}{{Massive Helicity-Chirality Spinor Formalism from
  Massless Amplitudes with On-shell Mass Insertion}}} (\bibinfo{year}{2025}),
  \eprint{2501.09062}.

\bibitem[{\citenamefont{Ema et~al.}(2024)\citenamefont{Ema, Gao, Ke, Liu, Lyu,
  and Mahbub}}]{Ema:2024vww}
\bibinfo{author}{\bibfnamefont{Y.}~\bibnamefont{Ema}},
  \bibinfo{author}{\bibfnamefont{T.}~\bibnamefont{Gao}},
  \bibinfo{author}{\bibfnamefont{W.}~\bibnamefont{Ke}},
  \bibinfo{author}{\bibfnamefont{Z.}~\bibnamefont{Liu}},
  \bibinfo{author}{\bibfnamefont{K.-F.} \bibnamefont{Lyu}}, \bibnamefont{and}
  \bibinfo{author}{\bibfnamefont{I.}~\bibnamefont{Mahbub}},
  \emph{\bibinfo{title}{{Momentum shift and on-shell constructible massive
  amplitudes}}}, \bibinfo{journal}{Phys. Rev. D}
  \textbf{\bibinfo{volume}{110}}, \bibinfo{pages}{105003}
  (\bibinfo{year}{2024}), \eprint{2403.15538}.

\bibitem[{\citenamefont{Ballav and Manna}(2021)}]{Ballav_2021}
\bibinfo{author}{\bibfnamefont{S.}~\bibnamefont{Ballav}} \bibnamefont{and}
  \bibinfo{author}{\bibfnamefont{A.}~\bibnamefont{Manna}},
  \emph{\bibinfo{title}{Recursion relations for scattering amplitudes with
  massive particles}}, \bibinfo{journal}{Journal of High Energy Physics}
  \textbf{\bibinfo{volume}{2021}} (\bibinfo{year}{2021}), ISSN
  \bibinfo{issn}{1029-8479},
  \urlprefix\url{http://dx.doi.org/10.1007/JHEP03(2021)295}.

\bibitem[{\citenamefont{Hill and Simmons}(2003)}]{Hill:2002ap}
\bibinfo{author}{\bibfnamefont{C.~T.} \bibnamefont{Hill}} \bibnamefont{and}
  \bibinfo{author}{\bibfnamefont{E.~H.} \bibnamefont{Simmons}},
  \emph{\bibinfo{title}{{Strong Dynamics and Electroweak Symmetry Breaking}}},
  \bibinfo{journal}{Phys. Rept.} \textbf{\bibinfo{volume}{381}},
  \bibinfo{pages}{235} (\bibinfo{year}{2003}), \bibinfo{note}{[Erratum:
  Phys.Rept. 390, 553--554 (2004)]}, \eprint{hep-ph/0203079}.

\bibitem[{\citenamefont{Weinberg}(1976)}]{Weinberg:1975gm}
\bibinfo{author}{\bibfnamefont{S.}~\bibnamefont{Weinberg}},
  \emph{\bibinfo{title}{{Implications of Dynamical Symmetry Breaking}}},
  \bibinfo{journal}{Phys. Rev. D} \textbf{\bibinfo{volume}{13}},
  \bibinfo{pages}{974} (\bibinfo{year}{1976}), \bibinfo{note}{[Addendum:
  Phys.Rev.D 19, 1277--1280 (1979)]}.

\bibitem[{\citenamefont{Susskind}(1979)}]{Susskind:1978ms}
\bibinfo{author}{\bibfnamefont{L.}~\bibnamefont{Susskind}},
  \emph{\bibinfo{title}{{Dynamics of Spontaneous Symmetry Breaking in the
  Weinberg-Salam Theory}}}, \bibinfo{journal}{Phys. Rev. D}
  \textbf{\bibinfo{volume}{20}}, \bibinfo{pages}{2619} (\bibinfo{year}{1979}).

\bibitem[{\citenamefont{Kaplan et~al.}(1984)\citenamefont{Kaplan, Georgi, and
  Dimopoulos}}]{Kaplan:1983sm}
\bibinfo{author}{\bibfnamefont{D.~B.} \bibnamefont{Kaplan}},
  \bibinfo{author}{\bibfnamefont{H.}~\bibnamefont{Georgi}}, \bibnamefont{and}
  \bibinfo{author}{\bibfnamefont{S.}~\bibnamefont{Dimopoulos}},
  \emph{\bibinfo{title}{{Composite Higgs Scalars}}}, \bibinfo{journal}{Phys.
  Lett. B} \textbf{\bibinfo{volume}{136}}, \bibinfo{pages}{187}
  (\bibinfo{year}{1984}).

\bibitem[{\citenamefont{Dugan et~al.}(1985)\citenamefont{Dugan, Georgi, and
  Kaplan}}]{Dugan:1984hq}
\bibinfo{author}{\bibfnamefont{M.~J.} \bibnamefont{Dugan}},
  \bibinfo{author}{\bibfnamefont{H.}~\bibnamefont{Georgi}}, \bibnamefont{and}
  \bibinfo{author}{\bibfnamefont{D.~B.} \bibnamefont{Kaplan}},
  \emph{\bibinfo{title}{{Anatomy of a Composite Higgs Model}}},
  \bibinfo{journal}{Nucl. Phys. B} \textbf{\bibinfo{volume}{254}},
  \bibinfo{pages}{299} (\bibinfo{year}{1985}).

\bibitem[{\citenamefont{Miransky
  et~al.}(1989{\natexlab{a}})\citenamefont{Miransky, Tanabashi, and
  Yamawaki}}]{Miransky:1988xi}
\bibinfo{author}{\bibfnamefont{V.~A.} \bibnamefont{Miransky}},
  \bibinfo{author}{\bibfnamefont{M.}~\bibnamefont{Tanabashi}},
  \bibnamefont{and} \bibinfo{author}{\bibfnamefont{K.}~\bibnamefont{Yamawaki}},
  \emph{\bibinfo{title}{{Dynamical Electroweak Symmetry Breaking with Large
  Anomalous Dimension and t Quark Condensate}}}, \bibinfo{journal}{Phys. Lett.
  B} \textbf{\bibinfo{volume}{221}}, \bibinfo{pages}{177}
  (\bibinfo{year}{1989}{\natexlab{a}}).

\bibitem[{\citenamefont{Miransky
  et~al.}(1989{\natexlab{b}})\citenamefont{Miransky, Tanabashi, and
  Yamawaki}}]{Miransky:1989ds}
\bibinfo{author}{\bibfnamefont{V.~A.} \bibnamefont{Miransky}},
  \bibinfo{author}{\bibfnamefont{M.}~\bibnamefont{Tanabashi}},
  \bibnamefont{and} \bibinfo{author}{\bibfnamefont{K.}~\bibnamefont{Yamawaki}},
  \emph{\bibinfo{title}{{Is the t Quark Responsible for the Mass of W and Z
  Bosons?}}}, \bibinfo{journal}{Mod. Phys. Lett. A}
  \textbf{\bibinfo{volume}{4}}, \bibinfo{pages}{1043}
  (\bibinfo{year}{1989}{\natexlab{b}}).

\bibitem[{\citenamefont{RUEGG and RUIZ-ALTABA}(2004)}]{RUEGG_2004}
\bibinfo{author}{\bibfnamefont{H.}~\bibnamefont{RUEGG}} \bibnamefont{and}
  \bibinfo{author}{\bibfnamefont{M.}~\bibnamefont{RUIZ-ALTABA}},
  \emph{\bibinfo{title}{The stueckelberg field}},
  \bibinfo{journal}{International Journal of Modern Physics A}
  \textbf{\bibinfo{volume}{19}}, \bibinfo{pages}{3265–3347}
  (\bibinfo{year}{2004}), ISSN \bibinfo{issn}{1793-656X},
  \urlprefix\url{http://dx.doi.org/10.1142/S0217751X04019755}.

\bibitem[{\citenamefont{Chanowitz and Gaillard}(1985)}]{Chanowitz:1985hj}
\bibinfo{author}{\bibfnamefont{M.~S.} \bibnamefont{Chanowitz}}
  \bibnamefont{and} \bibinfo{author}{\bibfnamefont{M.~K.}
  \bibnamefont{Gaillard}}, \emph{\bibinfo{title}{{The TeV Physics of Strongly
  Interacting W's and Z's}}}, \bibinfo{journal}{Nucl. Phys. B}
  \textbf{\bibinfo{volume}{261}}, \bibinfo{pages}{379} (\bibinfo{year}{1985}).

\bibitem[{\citenamefont{Gounaris et~al.}(1986)\citenamefont{Gounaris, Kogerler,
  and Neufeld}}]{Gounaris:1986cr}
\bibinfo{author}{\bibfnamefont{G.~J.} \bibnamefont{Gounaris}},
  \bibinfo{author}{\bibfnamefont{R.}~\bibnamefont{Kogerler}}, \bibnamefont{and}
  \bibinfo{author}{\bibfnamefont{H.}~\bibnamefont{Neufeld}},
  \emph{\bibinfo{title}{{Relationship Between Longitudinally Polarized Vector
  Bosons and their Unphysical Scalar Partners}}}, \bibinfo{journal}{Phys. Rev.
  D} \textbf{\bibinfo{volume}{34}}, \bibinfo{pages}{3257}
  (\bibinfo{year}{1986}).

\bibitem[{\citenamefont{Bagger and Schmidt}(1990)}]{Bagger:1989fc}
\bibinfo{author}{\bibfnamefont{J.}~\bibnamefont{Bagger}} \bibnamefont{and}
  \bibinfo{author}{\bibfnamefont{C.}~\bibnamefont{Schmidt}},
  \emph{\bibinfo{title}{{Equivalence Theorem Redux}}}, \bibinfo{journal}{Phys.
  Rev. D} \textbf{\bibinfo{volume}{41}}, \bibinfo{pages}{264}
  (\bibinfo{year}{1990}).

\bibitem[{\citenamefont{Chen}(2025)}]{Chen2025wig}
\bibinfo{author}{\bibfnamefont{J.}~\bibnamefont{Chen}},
  \emph{\bibinfo{title}{{Relativistic Particle on Light-Front}}}
  (\bibinfo{year}{2025}), \eprint{arxiv: 2510.08983}.

\bibitem[{\citenamefont{Chen et~al.}(2023)\citenamefont{Chen, Hagiwara,
  Kanzaki, and Mawatari}}]{Chen:2022gxv}
\bibinfo{author}{\bibfnamefont{J.}~\bibnamefont{Chen}},
  \bibinfo{author}{\bibfnamefont{K.}~\bibnamefont{Hagiwara}},
  \bibinfo{author}{\bibfnamefont{J.}~\bibnamefont{Kanzaki}}, \bibnamefont{and}
  \bibinfo{author}{\bibfnamefont{K.}~\bibnamefont{Mawatari}},
  \emph{\bibinfo{title}{{Helicity amplitudes without gauge cancellation for
  electroweak processes}}}, \bibinfo{journal}{Eur. Phys. J. C}
  \textbf{\bibinfo{volume}{83}}, \bibinfo{pages}{922} (\bibinfo{year}{2023}),
  \bibinfo{note}{[Erratum: Eur.Phys.J.C 84, 97 (2024)]}, \eprint{2203.10440}.

\end{thebibliography}
\bibliographystyle{apsper}

\end{document}